\documentclass[amsmath,amssymb,showpacs,twocolumn,superscriptaddress,pr]{revtex4-1}
 \usepackage{graphicx,bm,color,subfigure}
 \usepackage[latin9]{inputenc}
 \usepackage{colortbl}
 \usepackage{xcolor}
 \usepackage{bm}
 \usepackage{amsmath}
 \usepackage{hyperref}
 \hypersetup{colorlinks,
 	citecolor=blue}
 \usepackage{txfonts}
 \usepackage{epstopdf}
 \begin{document}
 	
 	\title{Lifshitz transition enhanced triplet $p_z$-wave superconductivity in hydrogen doped KCr$_3$As$_3$}
 	
 	\author{Ming Zhang}
 	\thanks{These two authors contributed equally to this work.}
 	\affiliation{School of Physics, Beijing Institute of Technology, Beijing 100081, China}

 	\author{Juan-Juan Hao}
 	\thanks{These two authors contributed equally to this work.}
 	\affiliation{School of Physics, Beijing Institute of Technology, Beijing 100081, China}
 	
 	\author{Xianxin Wu}
 	\email{xianxinwu@gmail.com}
 	\affiliation{Max-Planck-Institut f\"ur Festk\"orperforschung, Heisenbergstrasse 1, D-70569 Stuttgart, Germany}
 \affiliation{CAS Key Laboratory of Theoretical Physics, Institute of Theoretical Physics, Chinese Academy of Sciences, Beijing 100190, China}
 	
 	\author{Fan Yang}
 	\email{yangfan\_blg@bit.edu.cn}
 	\affiliation{School of Physics, Beijing Institute of Technology, Beijing 100081, China}

 	\begin{abstract}
    The recently synthesized air-insensitive hydrogen doped KCr$_3$As$_3$ superconductor has aroused great research interests. This material has, for the first time in the research area of the quasi-one-dimensional Cr-based superconductivity (SC), realized a tunability through charge doping, which will potentially significantly push the development of this area. Here based on the band structure from first-principle calculations, we construct a six-band tight-binding (TB) model equipped with multi-orbital Hubbard interactions, and adopt the random-phase-approximation approach to study the hydrogen-doping dependence of the pairing symmetry and superconducting $T_c$. Under the rigid-band approximation, our pairing phase diagram is occupied by the triplet $p_z$-wave pairing through out the hydrogen-doping regime $x\in (0.4,1)$ in which SC has been experimentally detected. Remarkably, the $x$-dependence of $T_c$ shows a peak at the 3D-quasi-1D Lifshitz transition point, although the total density of state exhibits a dip there. The corresponding doping level is near the experimental estimation of the optimal doping level. A thorough investigation of the band structure reveals type-II van-Hove singularities (VHSs) in the $\gamma$ band, which favor the formation of the triplet SC. It turns out that the $\gamma$- Fermi surface (FS) comprises two flat quasi-1D FS sheets almost parallel to the $k_z=0$ plane and six almost perpendicular tube-like FS sheets, and the type-II VHSs just lies in the boundary between these two FS parts. Furthermore, the $\left|k_z\right|$ of the VH planes reaches the maximum near the Lifshitz-transition point, which pushes the $T_c$ of the $p_z$-wave SC to the maximum. Our results appeal more experimental access into this intriguing superconductor.
 	\end{abstract}

 	%\pacs{74.20.-z, 74.20.Rp, 74.25.Dw}
 	
 	%75.10.-b General theory and models of magnetic ordering
 	%         (see also 05.50.+q Lattice theory and statistics)
 	%75.10.Lp Band and itinerant models
 	%74.20.-z Theories and models of superconducting state
 	%74.20.Rp Pairing symmetries (other than s-wave)
 	%74.25.Dw Superconductivity phase diagrams

 	\maketitle
 	
 	%%%%%%%%%%%%%%%%%%%%%%%%%%%%%%%%%%%%%%%%%%%%%%%%%%%%%%%%%%%%%%%%%%%%%%%%%%%%%%%%%%%

 \section{Introduction}
 In recent years, the quasi- one-dimensional (1D)  superconductors family  A$_2$M$_3$As$_3$ (A = Na, K, Rb, and Cs; M = Cr and Mo) ~\cite{Bao:15,Tang:15a,Tang:15,Pang:15,Zhi:15,Yang:15,Adroja:15,Kong:15,Balakirev:15,Wang:15,Pang:16,Cao:17,Adroja:17,Taddei:17,Zhao:18,Mu:18a,Mu:18,Luo:19} with highest superconducting $T_c$ above 10 K~\cite{Mu:18} have attracted tremendous research interests. These compounds consist of alkali-metal-atoms-separated [(M$_3$As$_3$)$^{2-}$]$_{\infty}$ double-walled subnanotubes with the low-energy degrees of freedom dominated by the M-3d orbitals~\cite{Jiang:15,Hu:15}, which are proposed to be strongly-correlated~\cite{Wu:15,Zhang:16,Zhou:17,Miao:16,Dai:15,Zhi:15,Yang:15,Taddei:17,Wu:1507}, implying an electron--interaction-driven pairing mechanism. Various experiments~\cite{Bao:15,Tang:15a,Tang:15,Zhi:15,Yang:15,Pang:15,Balakirev:15,Adroja:15,Adroja:17,Cao:17,Luo:19} have revealed unconventional pairing feature of this superconductors family, with evidences suggesting the existence of line gap nodes~\cite{Tang:15a,Pang:15} and possible spin triplet pairing states~\cite{Bao:15,Tang:15a,Tang:15,Yang:15,Cao:17,Luo:19,Wang:18,Yang:21}. This family, however, have a serious draw back in that these materials are instable in the atmosphere, which hinders the widespread experimental studies on them. Furthermore, the lack of tunability through charge doping, another shortcoming, prevents the understanding of the nature of the electron correlations.

 Slightly after the synthesization of the A$_2$Cr$_3$As$_3$ (233) family, its air-insensitive cousin family A$_1$Cr$_3$As$_3$ (133) were  obtained by removing half of the A$^+$ ions through an ethanol bath at room temperature~\cite{Bao:15_133, Tang:15_133,Mu:17_133,Liu:17_133}.  The 133 family shares similar quasi-1D crystalline structure and low-energy degrees of freedom with the 233 family~\cite{Cao:15}. Initially, there exists obvious conflict on the ground state property of the 133 family. While the works reported in Ref.~\cite{Bao:15_133, Tang:15_133, Feng:19} suggest the 133 family to be nonsuperconducting with a spin-glass ground state, definite evidence for superconductivity (SC) has been identified in the work reported in Ref.~\cite{Mu:17_133,Liu:17_133}. This conflict was finally reconciled by the revelation using neutron and X-ray diffraction~\cite{Taddei:17a} that the hydrogen atoms intercalated in the material play the crucial role for the appearance of SC~\cite{Taddei:17a, Xiang:19,Xiang:20}. The difference between nonsuperconducting and superconducting A$_1$Cr$_3$As$_3$ samples mainly lies in the hydrogen concentration, i.e. the stoichiometric formula of both samples should be A$_1$H$_x$Cr$_3$As$_3$ but their $x$ are different. The density functional theory (DFT)-based first-principle calculations suggest that the main role of the doped hydrogen atoms is to donate electrons~\cite{Taddei:17a, Wu:19}, whose concentration is now experimentally tunable~\cite{Taddei:17a,Xiang:19,Xiang:20}. Therefore, the hydrogen concentration $x$ in the 133 family provides an effective way, i.e. charge doping, to tune the correlated quantum states. For example, while the samples with $x<0.35$ are found to be nonsuperconducting with a spin-glass ground state, SC emerges in the samples with higher $x$, with the optimal $x$ for SC roughly estimated to be within the range of $x\in(0.65,0.71)$ ~\cite{Taddei:17a}.

 \begin{figure*}
 	\includegraphics[width=5.5in]{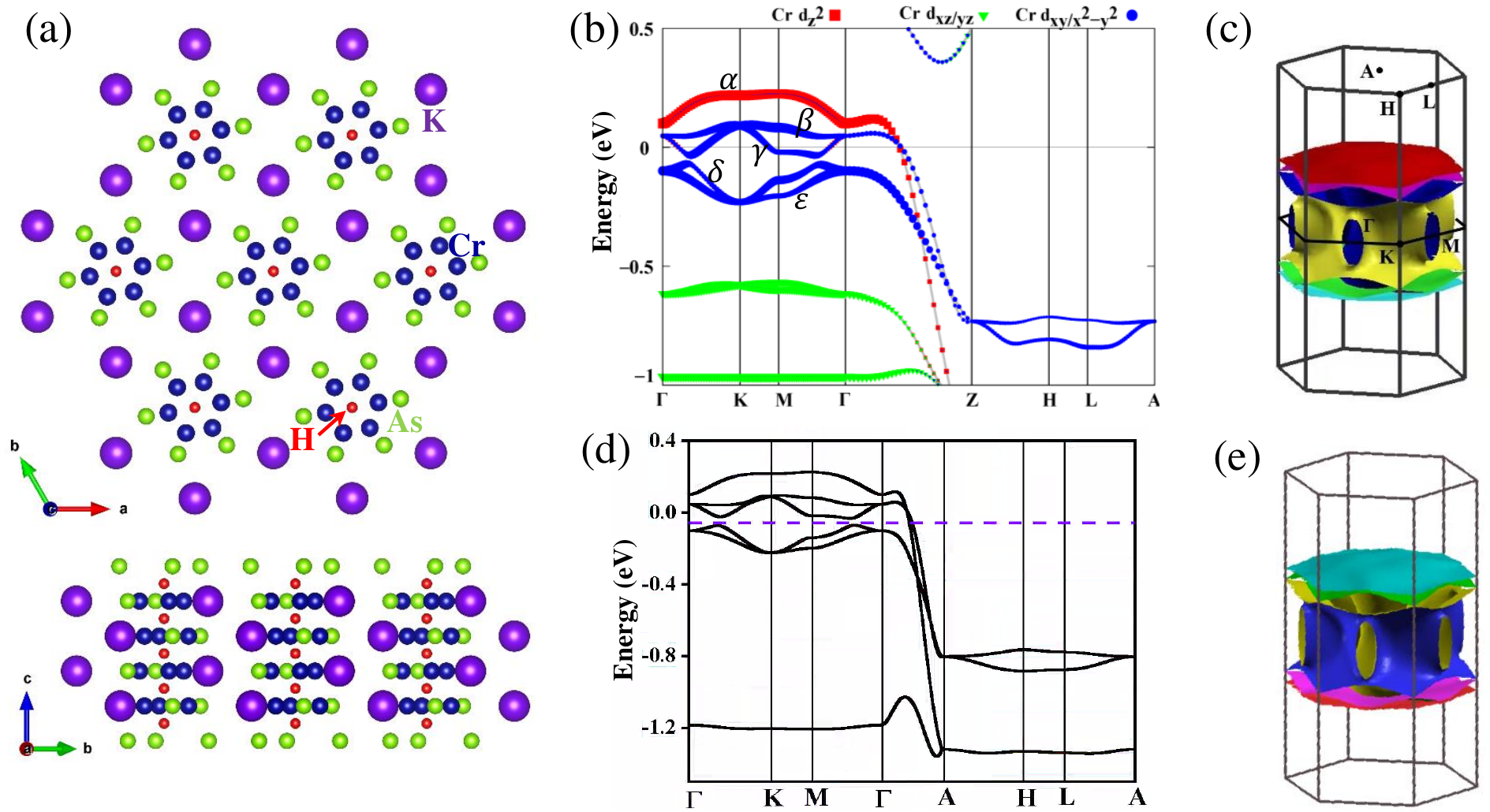}
 	\caption{(color online). (a) The top and side view for the crystal structure of KHCr$_3$As$_3$. (b) Band structure of KHCr$_3$As$_3$ along the high symmetry lines, with the red square /green triangle /blue circle being drown proportional to the weight of Cr-3d$_{z^2}$/-3d$_{xy}$/-3d$_{x^{2}-y^{2}}$, respectively. The five bands around the Fermi-level are marked as $\alpha$ - $\varepsilon$ respectively. (c) FSs of KHCr$_3$As$_3$ from DFT calculations, with the high symmetric points and the band indices for the three FSs marked. Band structures (d) and FSs (e) from the six-band TB model. Hydrogen doping at the purple dash is 0.6.}
 	\label{structure}
 \end{figure*}

 The DFT-based calculations~\cite{Taddei:17a} show that the chemical reaction between the KCr$_3$As$_3$ and the H$_2$ will form the KHCr$_3$As$_3$ with similar quasi-1D structure as that of KCr$_3$As$_3$, but with the hydrogen atoms intercalated at the center of Cr octahedra in the [(Cr$_3$As$_3$)$^{2-}$]$_{\infty}$ subnanotubes. No unstable phonon modes are found for this structure, suggesting its stability~\cite{Taddei:17a}. In the aspect of band structure~\cite{Taddei:17a, Wu:19}, the role of the intercalated hydrogen atoms mainly lie in the rise of the Fermi energy $E_F$, besides modest distortions to the bands near $E_F$. Therefore, we can say that in KHCr$_3$As$_3$, H has metallic bonding and acts as an electron donor. Furthermore the H concentration $x$ in the material is experimentally tunable~\cite{Taddei:17a, Wu:19}. While the DFT results for $x=0$ yield inter-layer-antiferromagnetic ordered ground state~\cite{Cao:15}, those for $x=1$ suggest non-magnetic ground state with short-ranged ferromagnetic~\cite{Wu:19} or antiferromagnetic~\cite{Taddei:17a} spin fluctuations, which might mediate SC. Therefore, the phase diagram in the KH$_x$Cr$_3$As$_3$ via tuning $x$ is like those of the cuprates and the iron-pnictide superconductors wherein magnetic order states are usually found to be proximate to the SC, suggesting the relevance of the e-e interaction driven pairing mechanism. However, detailed theoretical studies about this phase diagrams are still missing.

 A prominent feature of the band structure of KH$_x$Cr$_3$As$_3$ lies in the presence of a Lifshitz transition at about $x=0.75$~\cite{Wu:19}. From the DFT calculations, the low-energy degrees of freedom near $E_F$ for the KH$_x$Cr$_3$As$_3$ include the Cr-3d$_{xy}$, -3d$_{x^2-y^2}$ and -3d$_{z^2}$ orbitals. At $x=1$, there are three bands which cross the Fermi surface (FS), including the quasi-1D $\alpha$- and $\beta$- bands and the 3D $\gamma$- band. When the H concentration $x$ decreases to about $x=0.75$, the $\gamma$ band experiences a Lifshitz FS-topology transition, during which its 3D FSs are changed to two disconnected quasi-1D FS sheets (see Fig. 3 of Ref.~\cite{Wu:19}). The physical consequences of this Lifshitz transition, however, has not been thoroughly investigated.

 In this article, we study the pairing symmetry of the KH$_x$Cr$_3$As$_3$ via the random-phase-approximation (RPA) approach~\cite{RPA1,RPA2,RPA3,Kuroki2008,Scalapino2009,Scalapino2011,Liu2013,Liu2018,ZhangLD:19}, adopting the tight-binding (TB) model constructed from fitting our DFT band structure. Adopting the band structure for $x=1$, we use the rigid-band approximation to study the $x$- dependence of the pairing symmetry and the superconducting $T_c$ in the regime $x\in (0.4,1)$ wherein definite evidence of SC is experimentally detected~\cite{Taddei:17a}. Our results yield that the triplet $p_z$-wave pairing is the leading pairing symmetry in this doping regime. Particularly, the highest $T_c$ is obtained at the Lifshitz-transition doping level. A careful investigation of the band structure suggests that the presence of the type-II VHSs~\cite{VHS1,VHS2,VHS3,VHS4,VHS5} on the FSs are responsible for the triplet pairing, and the Lifshitz transition further favors the $p_z$-wave pairing symmetry. Our results appeal more experimental access into this intriguing superconductor hosting possible triplet $p_z$-wave topological SC.

 The rest of this paper is organized as follows: In Sec.~\ref{sec:DFT}, we provide our results from first-principle calculations based on DFT for the band structure of KHCr$_3$As$_3$, after which we construct its effective TB model. In Sec.~\ref{sec:phaseDiag}, we study the pairing symmetry of the system via the RPA approach, and present the pairing phase diagram. In Sec.~\ref{sec:Lifshitz}, we focus on the analysis of the band structure to reveal the role of the Lifshitz-transition and the type-II VHSs on the $\gamma$- FS, which favor the triplet $p_z$-wave SC. Our results are summarized in Sec.~\ref{sec:summary} together with some discussions about possible experimental implications.

 \section{Band structure and the TB Model}
 \label{sec:DFT}
 \subsection{The DFT band structure}
 \label{subsec:LDA}

 As shown in Fig.~\ref{structure} (a), the crystal structure of  KHCr$_3$As$_3$ is quasi-1D, and the basic unit is an infinite linear chain  double-walled sub-nanotubes (DWSN) [(Cr$_3$As$_3$)$^{2-}$]$_{\infty}$, which are connected to each other through K$^+$ alkaline cations~\cite{Taddei:17a}. Cr atoms should be covalently bonded with As atoms, and As atoms should be bonded with K$^+$ ions to separate the positively charged Cr and K atoms. They are composed of Cr$_6$ (or As$_6$) octahedrons on the shared surface along the crystallographic c direction and the H atom is located in the center of these octahedrons. The KHCr$_3$As$_3$ can be viewed as H-doped KCr$_3$As$_3$ with the doping level $x=1$. Similar to KCr$_3$As$_3$, KHCr$_3$As$_3$ has a centrosymmetric structure, with space group $P6_{3}/m$ (No.176)), (point group $C_{6h}$), in which Cr and As atoms form double-walled sub-nanotubes along the c axis.

%We have performed DFT calculations by using the Vienna {\it ab initio} calculation simulation package (VASP) \cite{Kresse1,Kresse2}. We employed the generalized gradient approximation (GGA) \cite{Perdew} in the form of Perdew-Burke-Ernzerhof (PBE) to describe the exchange-correlation effects. The core and valence electrons were treated within the Projector Augmented Wave (PAW) method \cite{Kresse3} with a cutoff of 400 eV for the plane wave basis. These calculations have been performed using a $4\times4\times10$ K-point grid, in the entire Brillouin zone. We adopt the lattice parameters $a = 9.0948$ \AA  ~and $c = 4.1770$ \AA  ~from Ref\cite{Taddei:17a}, which are obtained by lattice relaxation in the DFT calculations and are well consistent with experiment.
 The band structure of the KHCr$_3$As$_3$ material was calculated using the method of first-principles DFT theory as implemented in the QUANTUM ESPRESSO (QE) code ~\cite{Giannozzi}. The cutoff energy for expanding the wave functions into a plane-wave basis was
set to 60 Ry and the adopted K-point grid is $5\times5\times11$. The exchange correlation energy was described by the generalized gradient approximation (GGA) using the PBE functional~\cite{Perdew}.  The lattice parameters from relaxation are $a = 9.09481$ \AA~ and $c = 4.17703$ \AA , which are consistent with the experimental data in Ref.~\cite{Taddei:17a}. To obtain the six-band low-energy model, we initialize $d_{z^2}$ and $d_{xy/x^2-y^2}$ orbitals at the centers of Cr triangles and then perform the calculations of maximal localization for
these orbitals using Wannier90~\cite{Mostofi:14}.

Our band structure calculated from the DFT calculations is shown in Fig.~\ref{structure} (b) along the lines connecting the high-symmetry points marked in the Brillouin zone (BZ) shown in   Fig.~\ref{structure} (c). It can be seen that there are 5 bands near the Fermi level (marked as $\alpha, \beta, \gamma, \delta$ and $\epsilon$), among which only the three ones $\alpha$, $\beta$ and $\gamma$ cross the FS, which are mainly composed of 3d$_{z^2}$, 3d$_{x^{2}-y^{2}}$ and 3d$_{xy}$ orbitals of Cr atoms.  In comparison with the non-magnetic band structure of KCr$_3$As$_3$~\cite{Cao:15,ZhangLD:19}, our present one for the KHCr$_3$As$_3$ shows similar shape, with only modest distortion near the Fermi level that is relatively lift up by about 0.14 eV. Therefore, the inserted hydrogen atoms in the KHCr$_3$As$_3$ can be well viewed as effective electron donors, consistent with previous results ~\cite{Taddei:17a,Wu:19}. The FSs of the system are shown in Fig.~\ref{structure} (c), which include two quasi-1D FSs $\alpha$ and $\beta$, and one 3D FS $\gamma$. While the $\alpha$- and $\beta$- FSs each contains two disconnected FS sheets nearly parallel with the $xy$-plane, the $\gamma$- FS only contains one globally connected sheet.
   		\begin{figure}[htbp]
 	\centering
 	\includegraphics[width=0.45\textwidth]{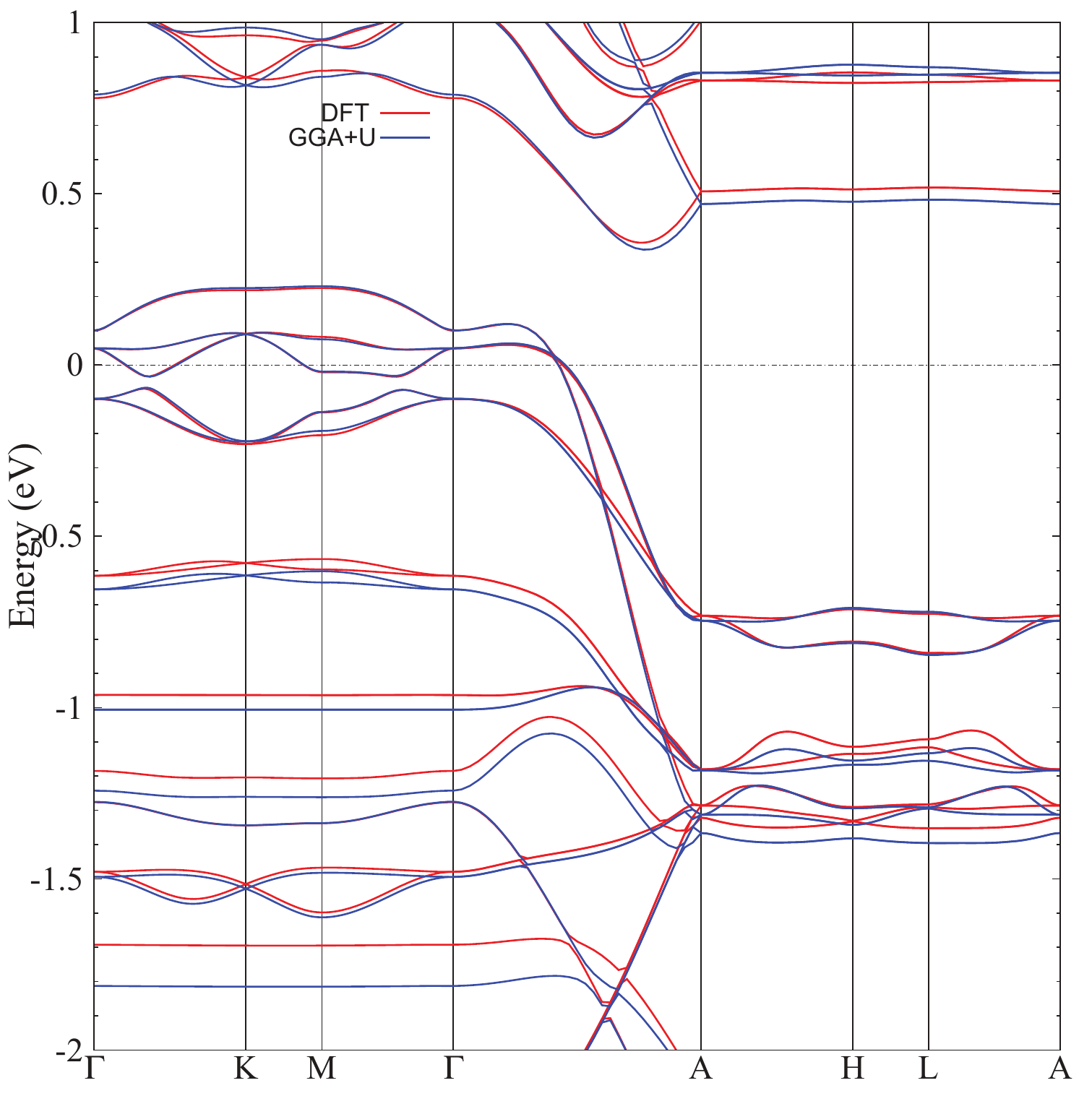}
 	\caption{(color online). Band structures of KHxCr3As3 with GGA (red lines) and GGA+U (blue lines,  U=2.3 eV and J=0.96 eV).}
 	\label{band+u}
 \end{figure}

  When the interaction in partially filled $d$ orbital is strong, an additional onsite interaction should be added in the calculations (GGA+U calculations) in order to get more accurate electronic structure. From available experimental evidence for KH$_x$Cr$_3$As$_3$, however, no strongly correlated state is clearly identified and thus the interaction is expected to not that strong. However, in check the robustness of the electronic structure, we performed GGA+U calculations with U=2.3 eV and J=0.96 eV (parameters from \cite{Mazin}) and the obtained band structure is displayed in Fig.~\ref{band+u}, in comparison with normal GGA calculations. We find that band structure from GGA+U calculations just changes slightly near the Fermi level and exhibit more noticeable change away from the Fermi level. Therefore the change of band structure at low energy is very small with including additional interactions. 	

In this paper, we neglect the spin-orbit coupling (SOC), as the atoms are not heavy in KHCr$_3$As$_3$. Including the SOC will introduce some gap opening around the $\Gamma$- and $K$- points\cite{Juanhao2022}. However, the relatively weak SOC will not change the band structure that much and thus will not change the pairing symmetry. Therefore, we focus on the band structure without SOC here.

 	\subsection{The TB Model}
 	\label{subsec:TB}
 	
As the bands near the Fermi level are predominantly contributed by Cr $d_{z^2}$, $d_{xy}$ and $d_{x^2-y^2}$ orbitals, we construct a six-band TB model to capture the low-energy bands in the DFT calculations, where $A_{1g}$ and $E_{2g}$ orbitals are located at the centers of two Cr triangles. This effective model is analogous to that of K$_2$Cr$_3$As$_3$ but with higher-symmetry point group $C_{6h}$~\cite{Wu:15}. To obtain the effective hopping parameters directly from DFT calculations, we initialize $d_{z^2}$ ($A_{1g}$) and $d_{xy/x^2-y^2}$ ($E_{2g}$) orbitals at the centers of Cr triangles [ (0,0,0/0.5) ] and then perform the calculations of maximal localization for these orbitals using Wannier90~\cite{Mostofi:14}. As the crystal symmetry is slightly broken in the resulted model, we further performed symmetrization on the obtained hopping parameters in real space using symmetry operations in $C_{6h}$. The obtained TB Hamiltonian in the momentum space which can be expressed as,
 	\begin{align}
 	H_{{\rm TB}}
 	=\sum_{\bm{k}\mu\nu\sigma}h_{\mu\nu}(\bm{k})
 	c^{\dagger}_{\mu\sigma}(\bm{k})c_{\nu\sigma}(\bm{k}),
 	\end{align}
 	Here $\mu,\nu=1,\cdots, 6$ indicating the orbital-sublattice indices, containing the $d_{z^2}$, $d_{x^2-y^2}$ and $d_{xy}$ orbitals of A sublattice and B sublattice. The elements of the $h(\bm{k})$ matrix is,
 	\begin{align}
 	h_{\mu\nu}(\bm{k})=\sum_{r_1,r_2,r_3}t^{r_1,r_2,r_3}_{\mu\nu}e^{i\bm{k}\cdot(r_1\bm{a_1}+r_2\bm{a_2}+r_3\bm{a_3})}.\label{TB_parameters}
 	\end{align}
 	with $\bm{a_1}=(\frac{\sqrt{3}}{2}a_0,-\frac{1}{2}a_0,0)$, $\bm{a_2}=(0,a_0,0)$ and $\bm{a_3}=(0,0,c_0)$. The data of $t^{r_1,r_2,r_3}_{\mu\nu}$ for $r_{1}\in[-4,4]$, $r_2\in[-2,2]$, $r_3\in[-6,6]$ and $\mu,\nu\in[1,6]$ is provided in the Supplementary Material (SM)~\cite{SuppMat}. Note that in the absence of SOC and magnetic order, the time-reversal symmetry requires these hopping parameters to be real. The band structure from this model is shown in Fig.~\ref{structure} (d), which in good agreement with that of the DFT (Fig.~\ref{structure} (b)) near the Fermi level.

 Although the above provided band structure and TB model only accurately adapt to the KHCr$_3$As$_3$, we take the rigid-band approximation and adopt them to describe the band structure of KH$_x$Cr$_3$As$_3$ with only the chemical potential tuned according to the variation of $x$. Note that each unit cell contains two H and each H donates one electron. The validity of this approximation is based on the similarity between the band structures of KHCr$_3$As$_3$ ($x=1$) and KCr$_3$As$_3$ ($x=0$)\cite{Taddei:17a}. However, since the two band structures are not exactly the same, we limit our study to the regime not too far from $x=1$ where the rigid-band approximation adapts better. In our calculations, we set $x$ to be within the doping regime $x\in (0.4,1)$, in which definite evidence of SC have been detected~\cite{Taddei:17a, Wu:19}.

 	\section{The RPA-based pairing phase-diagram}
 	\label{sec:phaseDiag}
 	We adopt the following extended Hubbard model Hamiltonian in our study:
 	\begin{align}\label{model}
 	H=&H_{\text{TB}}+H_{int}\nonumber\\
 	H_{int}=&U\sum_{i\mu}n_{i\mu\uparrow}n_{i\mu\downarrow}+
 	V\sum_{i,\mu<\nu}n_{i\mu}n_{i\nu}+J_{H}\sum_{i,\mu<\nu}                   \nonumber\\
 	&\Big[\sum_{\sigma\sigma^{\prime}}c^{+}_{i\mu\sigma}c^{+}_{i\nu\sigma^{\prime}}
 	c_{i\mu\sigma^{\prime}}c_{i\nu\sigma}+(c^{+}_{i\mu\uparrow}c^{+}_{i\mu\downarrow}
 	c_{i\nu\downarrow}c_{i\nu\uparrow}+h.c.)\Big]
 	\end{align}
 	Here, $n_{i\mu}=n_{i\mu\uparrow}+n_{i\mu\downarrow}$ denotes that the number of electrons in orbital $\mu$ at lattice site $i$. $c^{+}_{i\mu\sigma}(c_{i\mu\sigma})$ is the electron creation (annihilation) operator at lattice site $i$ with orbital $\mu$ and spin $\sigma$. The interaction parameters $U$, $V$, and $J_H$ denote the intra-orbital, inter-orbital Hubbard repulsion, and the Hund's rule coupling (as well as the pair hopping) respectively, which satisfy the relation $U=V+2J_H$.

 	\subsection{Bare Susceptibility}
 	\label{sec:Susceptibility}
 	
  		\begin{figure}[htbp]
 		\centering
 		\includegraphics[width=0.5\textwidth]{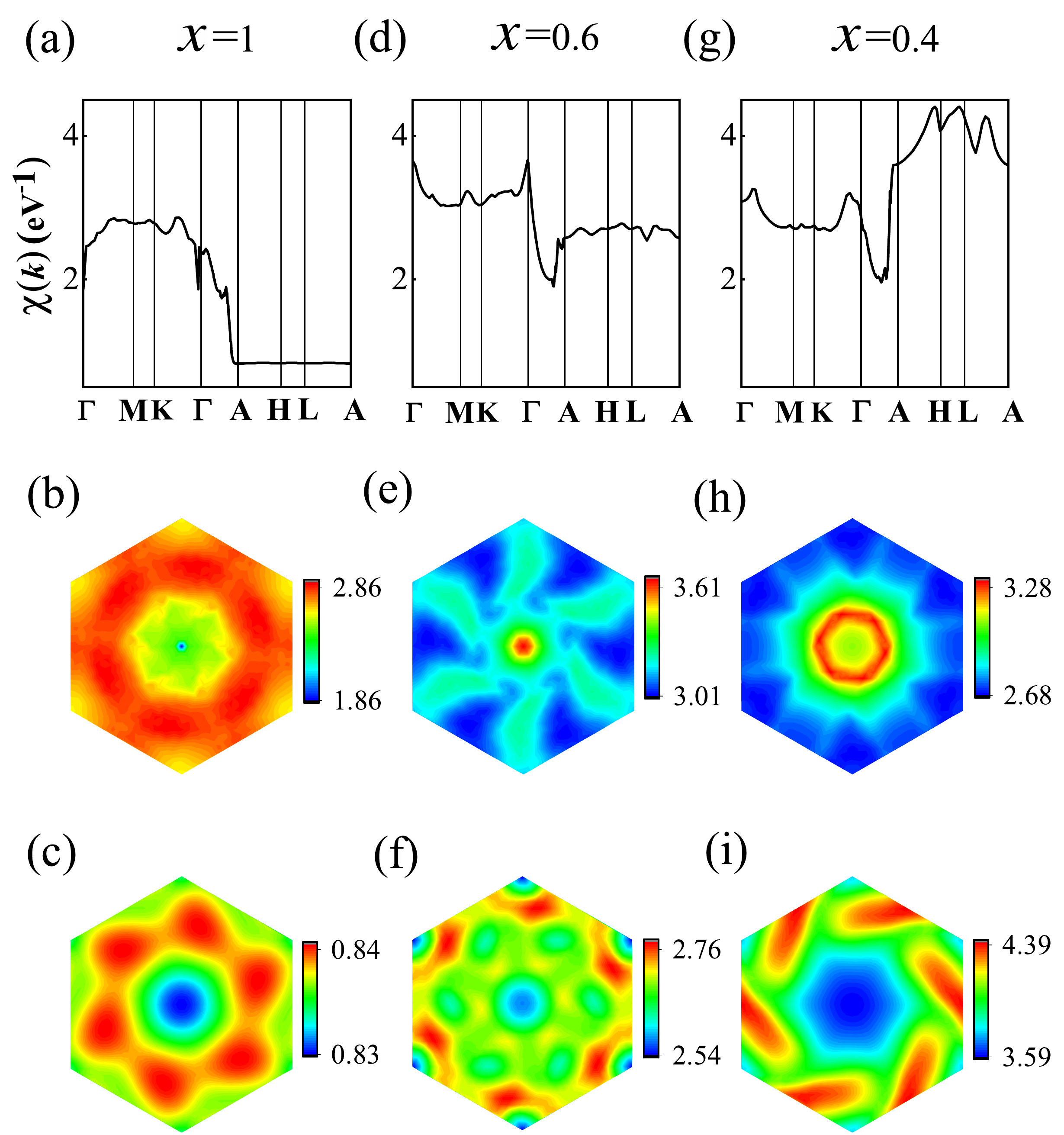}
 		\caption{(color online). The $\bm{k}$-space distribution of the largest eigenvalue $\chi(\bm{k})$ of the susceptibility matrix $\chi^{(0)pq}_{st}(\bm{k},i\omega_n=0)$ for (a)-(c) KCr$_3$As$_3$H, (d)-(f) KCr$_3$As$_3$H$_{0.6}$ and (g)-(i) KCr$_3$As$_3$H$_{0.4}$. From top to bottom are largest eigenvalue of the susceptibility matrix $\chi^{(0)}_{pqst}(\bm{k},i\omega_n=0)$ along the high-symmetry lines in the Brillouin zone, on the $k_z=0$ plane and on the $k_z=\pi$ plane, respectively.}
 		\label{chi}
 	\end{figure}
 	 	
 We first define the following bare susceptibility tensor in the normal state for the non-interacting case:
 	\begin{align}\label{chi0}
 	\chi^{(0)}_{pqst}(\bm{k},\tau)\equiv
 	&\frac{1}{N}\sum_{\bm{k}_1\bm{k}_2}\left\langle
 	T_{\tau}c_{p}^{\dagger}(\bm{k}_1,\tau)
 	c_{q}(\bm{k}_1+\bm{k},\tau)\right.                      \nonumber\\
 	&\left.\times c_{s}^{\dagger}(\bm{k}_2+\bm{k},0)
 	c_{t}(\bm{k}_2,0)\right\rangle_0,
 	\end{align}
 	Here $\langle\cdots\rangle_0$ denotes the thermal average for the noninteracting system, $T_{\tau}$ denotes the imaginary time-ordered product, and the tensor indices $p,q,s,t=1,\cdots,6$ denote the orbital-sublattice indices. Fourier transformed to the imaginary frequency space, the bare susceptibility can be expressed by the following explicit formulism:
 	\begin{align}\label{chi0e}
 	\chi^{(0)}_{pqst}(\bm{k},i\omega_n)
 	=&\frac{1}{N}\sum_{\bm{k}'\alpha\beta}
 	\xi^{\alpha}_{t}(\bm{k}')
 	\xi^{\alpha*}_{p}(\bm{k}')
 	\xi^{\beta}_{q}(\bm{k}'+\bm{k})                         \nonumber\\
 	&\times\xi^{\beta*}_{s}(\bm{k}'+\bm{k})
 	\frac{n_F(\varepsilon^{\beta}_{\bm{k}'+\bm{k}})
 		-n_F(\varepsilon^{\alpha}_{\bm{k}'})}
 	{i\omega_n+\varepsilon^{\alpha}_{\bm{k}'}
 		-\varepsilon^{\beta}_{\bm{k}'+\bm{k}}}.
 	\end{align}
 	where $\alpha,\beta=1,\cdots,6$ are band indices, $\varepsilon^{\alpha}_{\bm{k}}$ and $\xi^{\alpha}\left(\bm{k}\right)$ are the $\alpha$-th eigenvalue (relative to the chemical potential $\mu_c$) and eigenvector of the TB model, respectively, and $n_F$ is the Fermi-Dirac distribution function.

 The susceptibility tensor $\chi^{(0)}_{pqst}(\bm{k},i\omega)$ defined on the above can be viewed as a matrix $\chi^{(0)pq}_{st}(\bm{k},i\omega)$ by taking the combined $pq$ indices as the row index and the combined $st$ indices as the column index. In Fig.~\ref{chi}, we show the $\bm{k}$-dependence of the largest eigenvalue $\chi(\bm{k})$ of the zero-frequency susceptibility matrix $\chi^{(0)pq}_{st}(\bm{k},i\omega_n=0)$ for three different doping levels, i.e. $x=1$ in (a)-(c), $x=0.6$ in (d)-(f) and $x=0.4$ in (g)-(i). Among these figures, the (a), (d) and (g) in the first row are along the high-symmetry lines in the BZ; the (b), (e) and (h) in the second row are on the $k_z=0$ plane;  and the (c), (f) and (i) in the third row are on the $k_z=\pi$ plane. Note that here $x=1$ denotes KHCr$_3$As$_3$, $x=0.4$ is the lowest electron-doping level we consider, and the doping level $x=x_c=0.6$ indicates the Lifshitz-transition point in our TB model. This doping level is slightly lower than the $x_c=0.73$ in our DFT band structure obtained via the QE code and the $x_c=0.75$ in previous DFT band structure obtained via the VASP code~\cite{Wu:19}, mainly due to the slight deviation between our TB model and the DFT band structures.

Figure~\ref{chi} illustrates two doping-dependent features for the distributions of the susceptibilities in the BZ. The first feature lies in that the spin correlations are globally enhanced when the electron-doping level is decreased from $x=1$ (for the KHCr$_3$As$_3$) to $x=0$ (for the KCr$_3$As$_3$). For example, let's focus on the doping dependence of the maximum value of $\chi(\bm{k})$ through out the BZ, i.e. the peak value $\chi_{Max}$ at the momentum $k_0$ for a fixed doping level $x$. For $x=1$, $\chi_{Max}$ is about 2.9 and the corresponding $k_0$ is within the $k_z=0$ plane, as shown in Fig.~\ref{chi} (b); for $x=0.6$, $\chi_{Max}$ is about 3.6 around the $\Gamma$ point, as shown in Fig.~\ref{chi} (e); while for $x=0.4$, $\chi_{Max}$ is further enhanced to 4.4 and the corresponding $k_0$ moves to the $k_z=\pi$ plane, as shown in Fig.~\ref{chi} (i). This feature suggests that the tendency toward magnetic order increases from KHCr$_3$As$_3$ to KCr$_3$As$_3$, which is consistent with the experiments~\cite{Taddei:17a,Feng:19} and previous DFT results~\cite{Cao:15,Wu:19}. The second feature lies in that the momentum $\bm{k}_0$where the susceptibility peaks gradually shifts from within the $k_z=0$ plane to within the $k_z=\pi$ plane, reflecting the variation from inter-layer ferromagnetic correlations for KHCr$_3$As$_3$ to inter-layer antiferromagnetic correlations for KCr$_3$As$_3$, also consistent with previous DFT calculations~\cite{Cao:15,Wu:19}.

Although the spin fluctuations in both the $x=1$ and $x=0.6$ cases are inter-layer ferromagnetic, there is obvious difference between them in the aspect of intra-layer pattern. Figure ~\ref{chi} (a) and (b) show that the susceptibility for $x=1$ is smooth in the $k_z=0$- plane without obvious peaks. Therefore, the intra-layer spin fluctuation pattern for this doping level are neither typical ferromagnetic nor typical antiferromagnetic, but rather their competition, consistent with Ref.~\cite{Taddei:17a}. The situation changes for the Lifshitz-transition doping $x=0.6$, for which Fig.~\ref{chi} (d) and (e) show that the susceptibility sharply peaks at the $\Gamma$-point, implying typical ferromagnetic spin fluctuations. In Fig.~\ref{chi_sc_phase} (a), the doping dependence of the susceptibility for the $\Gamma$-point is shown, which exhibits a peak near $x=0.6$, suggesting that the ferromagnetic spin fluctuations are strongest near the Lifshitz transition. Such typical ferromagnetic fluctuations can favor the formation of spin-triplet SC, as will be shown in the following.

\subsection{The RPA approach}
\label{sec:RPA}
 	We further calculate the spin $(s)$ and charge $(c)$ susceptibilities 	following the standard multi-orbital RPA approach~\cite{RPA1,RPA2,RPA3,Kuroki2008,Scalapino2009,Scalapino2011,Liu2013,Liu2018,ZhangLD:19}, see also the Appendix. At the RPA level, the renormalized spin and charge susceptibilities of the system read
 \begin{align}\label{chisce}
 \chi^{(s,c)}(\bm{k},i\omega_n)=[I\mp\chi^{(0)}(\bm{k},i\omega_n)
 U^{(s,c)}]^{-1}\chi^{(0)}(\bm{k},i\omega_n),
 \end{align}
 Here the nonzero elements $U^{(s,c)\mu\nu}_{\theta\xi}$ of $U^{(s,c)}$ satisfy $\mu,\nu,\theta,\xi\leq 3$ or $>3$ simultaneously, which are as follow,
 \begin{align}
 U^{(s(c))\mu\nu}_{\theta\xi}=\left\{
 \begin{array}{ll}
 U(U),  & \mu=\nu=\theta=\xi; \\
 J_H(2V-J_H), & \mu=\nu\neq\theta=\xi; \\
 J_H(J_H), & \mu=\theta\neq\nu=\xi; \\
 V(2J_H-V),  & \mu=\xi\neq\theta=\nu.
 \end{array}
 \right.
 \end{align}
 In Eq.~(\ref{chisce}), $\chi^{(s,c,0)}(\bm{k},i\omega_n)$ and $U^{(s,c)}$ are operated as $6^{2}\times 6^{2}$ matrices (see for example in Ref.~\cite{Liu2013}).

 		\begin{table}
 		\centering
 		\caption{The twelve possible pairing symmetries for KCr$_3$As$_3$H in the absence of SOC, among which six are spin-singlet while the rest are spin-triplet.}
 		\label{Tab:one}
 		\begin{tabular}{@{}ccccccccccc@{}}
 			\\\hline\hline
 			singlet   &&&&&&&&&&  triplet   \\
 			\hline\hline
 			$s$          &&&&&&&&&&  $p_z$  \\
 			$(d_{x^2-y^2},d_{xy})$             &&&&&&&&&&      $(d_{x^2-y^2},d_{xy})\cdot p_z$             \\
 			$(p_x,p_y)\cdot p_z$         &&&&&&&&&&  $(p_x,p_y)$          \\
 			$f_{x^3-3xy^2}\cdot p_z$         &&&&&&&&&&  $f_{x^3-3xy^2}$             \\
            $f'_{y^3-3x^2y}\cdot p_z$         &&&&&&&&&&  $f'_{y^3-3x^2y}$             \\
            $f_{x^3-3xy^2}\cdot f'_{y^3-3x^2y}$         &&&&&&&&&&  $f_{x^3-3xy^2}\cdot f'_{y^3-3x^2y}\cdot p_z$             \\
 			\hline\hline
 		\end{tabular}
 	\end{table}

\begin{figure}[htbp]
	\centering
	\includegraphics[width=0.45\textwidth]{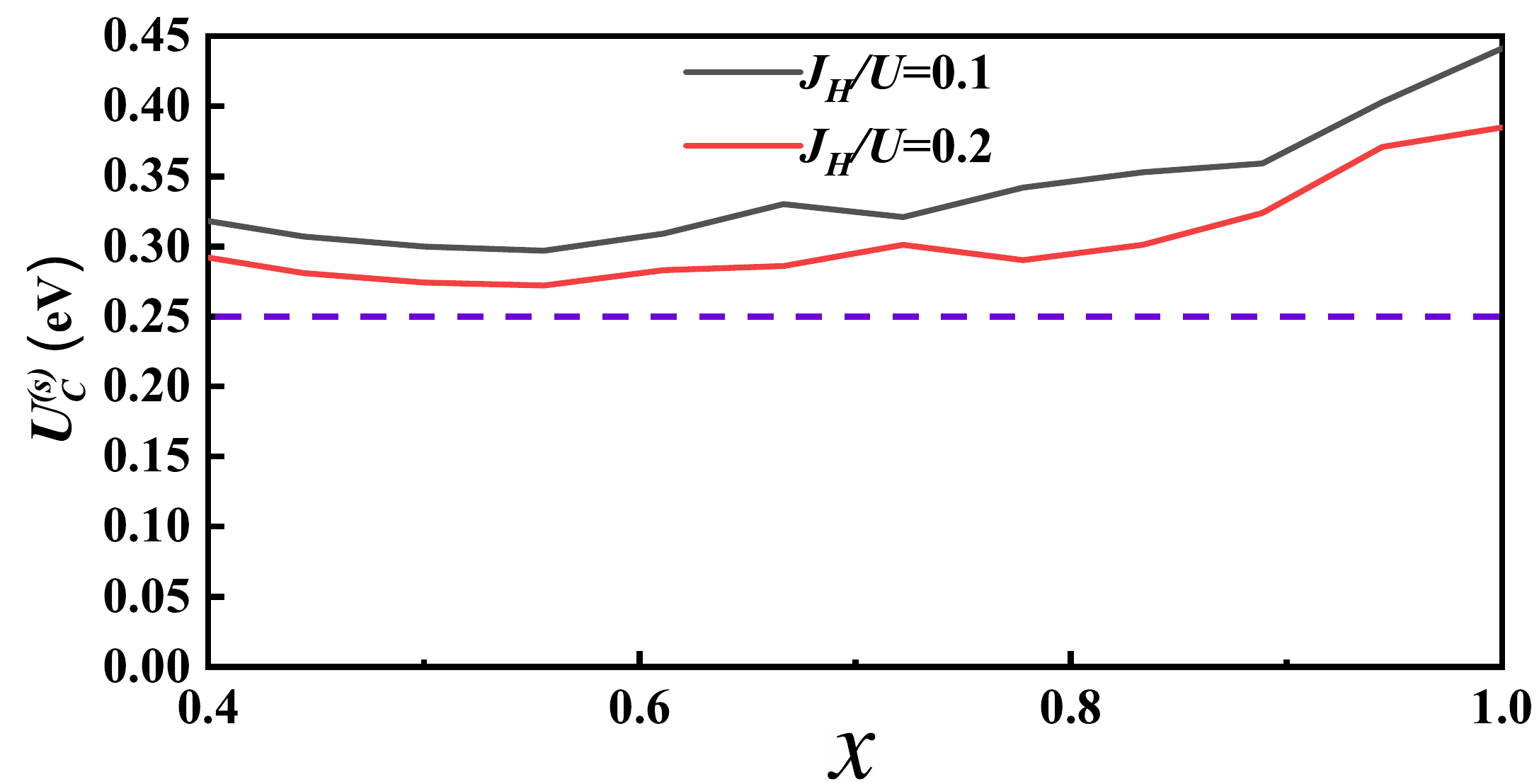}
	\caption{(color online). Critical interaction $U_c$ of the RPA spin susceptibility a function of hole doping. The red and gray lines are for $J_H$=0.2U and $J_H$=0.1U, respectively. In the main text, we adopt U=0.25 eV to calculate the pairing strength, in order to avoid magnetic instability.} 		\label{usc}
\end{figure}

Our numerical results suggest that the repulsive Hubbard interactions suppress the charge susceptibility, but enhance the spin susceptibility, consistent with previous results~\cite{RPA1,RPA2,RPA3,Kuroki2008,Scalapino2009,Scalapino2011,Liu2013,Liu2018,Kohn:65,Raghu:10,Cho:13,Scalapino2012}. There is a critical interaction strength $U_c$, where the spin susceptibility diverges, implying the formation of spin density wave (SDW). The doping dependences of $U_c$ for $J_H=0.1U$ and $J_H=0.2U$ are shown in Fig.~\ref{usc}. At $U<U_c$, Cooper pairing may develop through exchanging spin and/or charge fluctuations. In particular, we consider Cooper pair scatterings both within and between the bands, hence both intra- and inter-band effective interactions $V^{\alpha\beta}(\mathbf{k,k'})$~\cite{Wu:15} (here $\alpha/\beta=1,\cdots,6$ are band indices) are accounted for.  From the effective interaction vertex $V^{\alpha\beta}(\mathbf{k,k'})$, we obtain the following linearized gap equation near the superconducting $T_c$:
 	\begin{align}\label{gapeq}
 	-\frac{1}{(2\pi)^3}\sum_{\beta}\oiint_{FS}
 	d^{2}\bm{k}'_{\Vert}\frac{V^{\alpha\beta}(\bm{k},\bm{k}')}
 	{v^{\beta}_{F}(\bm{k}')}\Delta_{\beta}(\bm{k}')=\lambda
 	\Delta_{\alpha}(\bm{k}).
 	\end{align}
 Here the integration runs along the $\beta$- FS, the Fermi velocity $v^{\beta}_F(\bm{k}')$ is the amplitude of the gradient of the band energy at the momentum $\bm{k}'$, and $\bm{k}'_\parallel$ is the projection of $\bm{k}'$ on the FS. Superconducting pairing in various channels emerge as the eigenstates of the above gap equation. The leading pairing $\Delta_\alpha(\bm{k})$ is given by the eigenstate corresponding to the largest eigenvalue $\lambda$. The critical temperature $T_c$ is related to $\lambda$ through $T_c\propto e^{-1/\lambda}$.
 	
	The eigenvector(s) $\Delta_{\alpha}(\bm{k})$ for each eigenvalue $\lambda$ obtained from gap equation (\ref{gapeq}) as the basis function(s) forms an irreducible representation of the $C_{6h}$ point group. In the absence of SOC, twelve possible pairing symmetries are possible candidates for the system, which include six singlet pairings and six triplet pairings, as listed in Table~\ref{Tab:one}.

 	\subsection{The pairing phase diagram}
 	\label{sec:phase_diagram}

\begin{figure}[htbp]
	\centering
 	\includegraphics[width=0.48\textwidth]{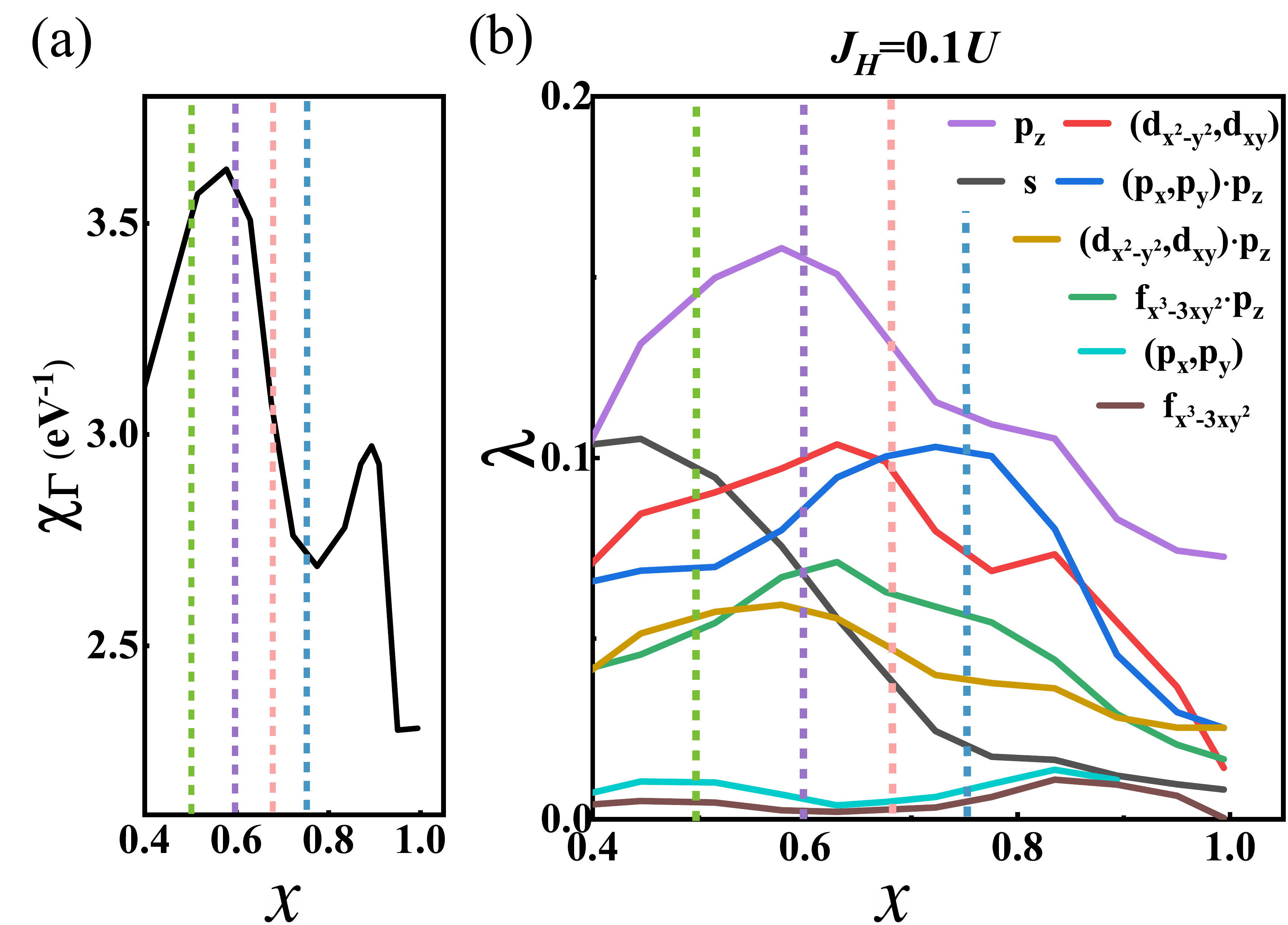}
 	\caption{(color online). (a) The hydrogen-doping level $x$- dependence of the susceptibility $\chi(\mathbf{k})$ for the $\Gamma$-point. (b) The largest pairing eigenvalues $\lambda$ as function of $x$ for eight pairing symmetries with relatively higher $\lambda$ under $U$ = 0.25eV, $J_H$ = 0.1$U$. The dotted lines with different colors mark different $x$ in KH$_x$Cr$_3$As$_3$. Specifically, the green, purple, pink and blue lines mark $x=0.5$, $x=0.6$, $x=0.67$ and $x=0.75$, respectively.} 		\label{chi_sc_phase}
 \end{figure}

  \begin{figure}[htbp]
	\centering
	\includegraphics[width=0.45\textwidth]{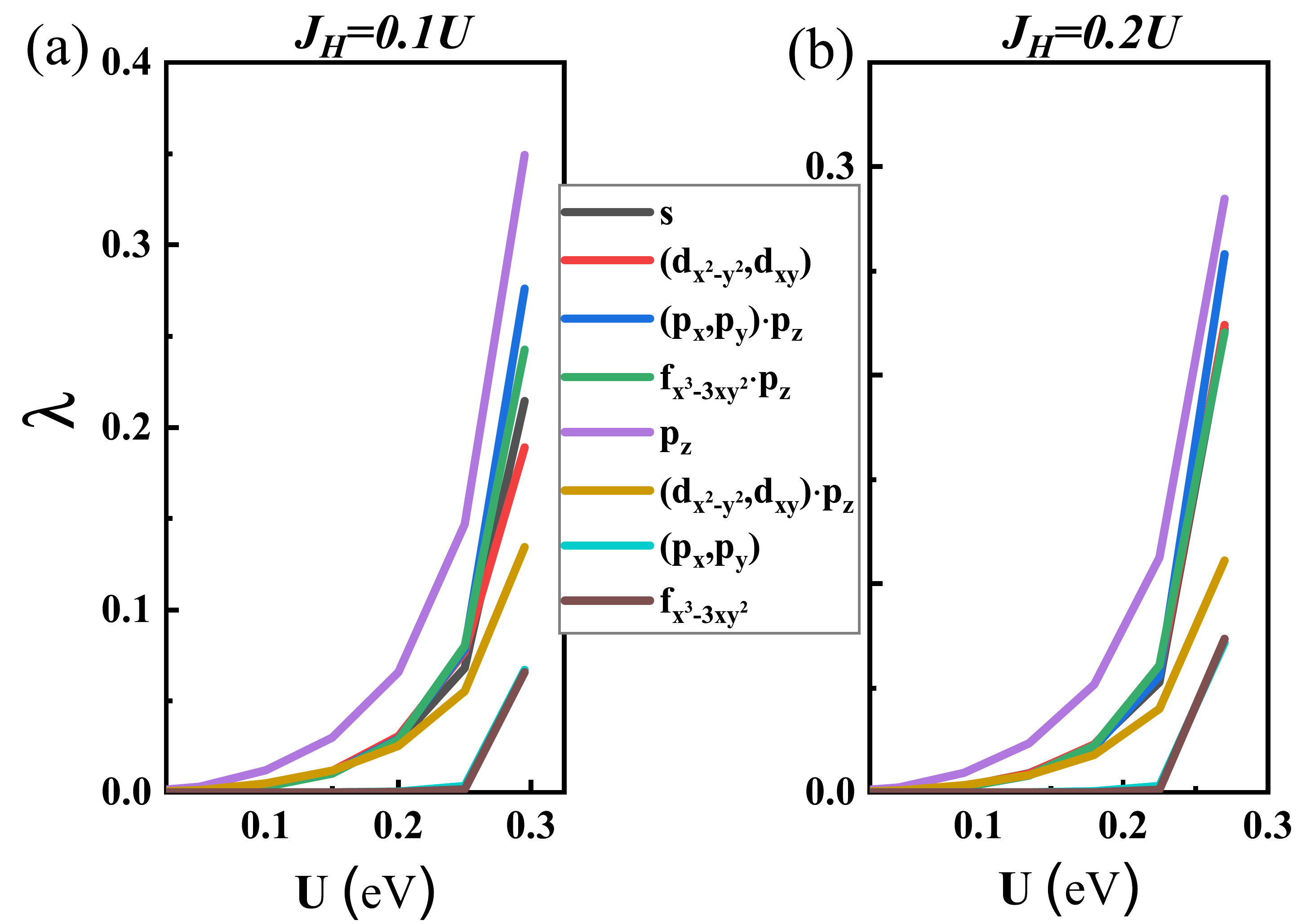}
	\caption{(color online). Leading pairing eigenvalue $\lambda$ as function of interaction $U$ at x=0.6 for (a) $J_H = 0.1U$ and (b) $J_H =0.2U$. Only the eight strongest pairing-symmetry channels are shown. }
	\label{phase1}
\end{figure}

The doping $x$ dependence of the largest pairing eigenvalues $\lambda$ for various pairing symmetries are shown in Fig.~\ref{chi_sc_phase} (b). The parameter settings are $J_H=0.1U$ and $U=0.25$ eV, satisfying $U<U_c$, as shown in Fig.~\ref{usc}. The $U$ dependence of $\lambda$ is shown in Fig.~\ref{phase1} for $J_H=0.1U$ in (a) and $J_H=0.2U$ in (b) at $x=0.6$. Eight out of the twelve possible pairing symmetries with relatively higher pairing eigenvalues are shown. Here we only consider the regime $x>0.4$ because of the following two reasons. On the one hand, too low $x$ might invalidate the rigid-band approximation as the band structure we adopt is for $x=1$. On the other hand, the spin-glass phase instead of SC is experimentally detected for $x<0.4$~\cite{Taddei:17a,Feng:19}, suggesting that the system should have already entered the spin-ordered phase in that doping regime, which invalidates the RPA treatment.

Two important results are provided by Fig.~\ref{chi_sc_phase} (b). Firstly, the triplet $p_z$-wave pairing is the leading pairing symmetry in the whole doping regime of $x\in (0.4,1)$ relevant to experiments. This result suggests that the SC detected by experiments should be of $p_z$-wave pairing symmetry. Secondly, the doping-dependence of the $\lambda$ and hence the $T_c$ of the obtained $p_z$-wave SC takes a domed shape peaking near the Lifshitz-transition point with $x\approx 0.6$. What's more, a comparison between Fig.~\ref{chi_sc_phase} (a) and (b) reveals the similarity between the $\lambda\sim x$ relation for the triplet $p_z$-wave SC and the $\chi_{\Gamma}\sim x$ relation. The physical reason for such similarity lies in that the ferromagnetic fluctuation reflected by $\chi_{\Gamma}$ favors the formation of triplet SC.

	\begin{figure}[htbp]
 	\centering
 	\includegraphics[width=0.45\textwidth]{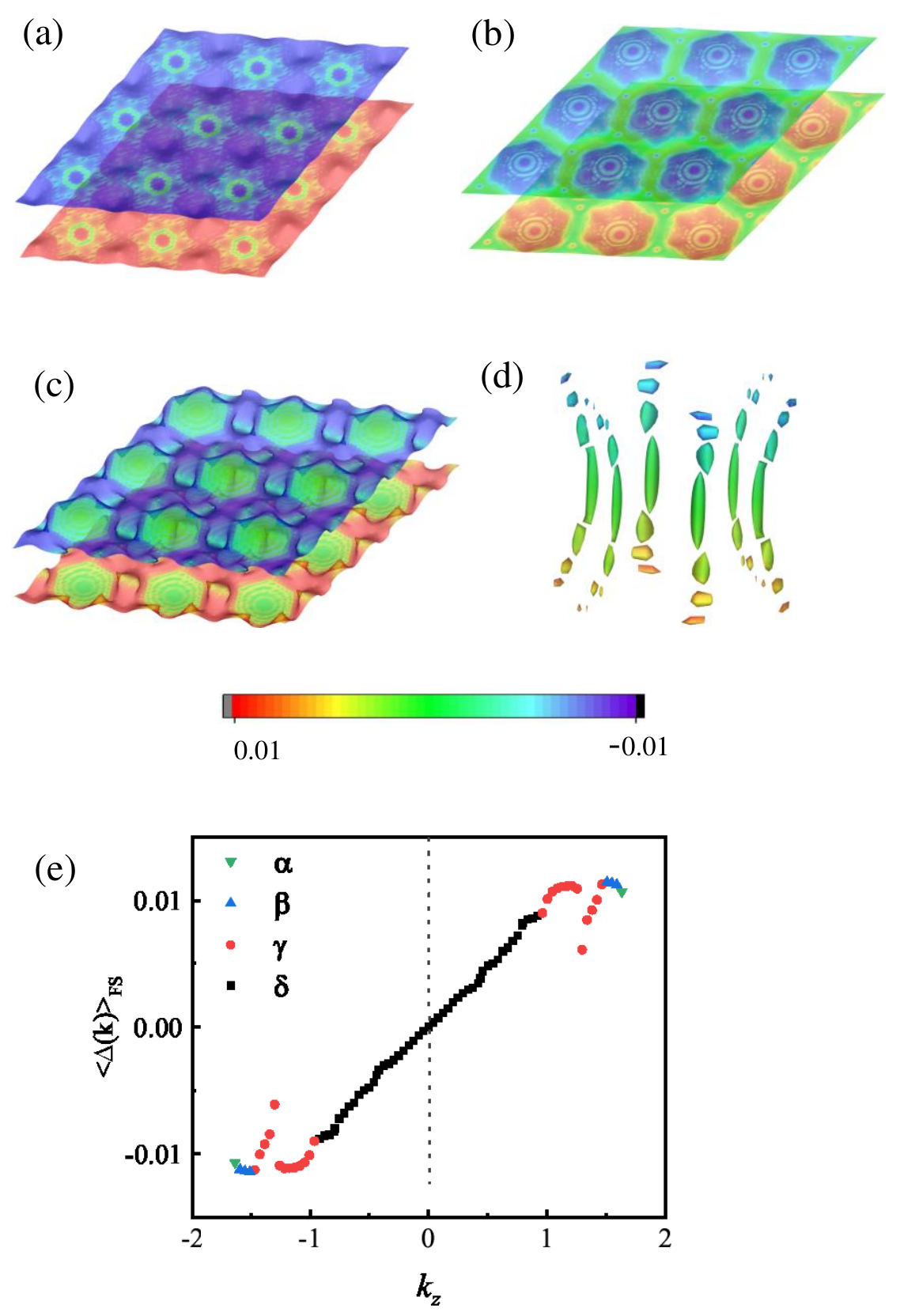}
  	\caption{(color online). (a) The pairing gap functions shown on different FSs for the $p_z$-SC with $x=0.6$. The FSs shown from (a) to (d) are $\alpha$, $\beta$, $\gamma$ and $\delta$ respectively. The three FSs have different $k_z$ scales, for $\alpha$-FS, $k_z$=$\pm 1.62$, for $\beta$-FS, $k_z$=$\pm 1.62$, for $\gamma$-FS, $k_z$=$\pm 1.45$ and for $\delta$-FS, $k_z$=$\pm 0.8$} (e) The $k_z$-dependence of the gap function averaged on the FS with fixed $k_z$.
 			\label{gap}
 		\end{figure}

The distribution of the relative gap function of the obtained $p_z$-wave SC is shown on the $\alpha$-, $\beta$- $\gamma$-, and $\delta$- FSs for the Lifshitz-transition doping level $x=0.6$ in Fig.~\ref{gap} (a) - (d). While the $\alpha$-, $\beta$- and $\gamma$- FSs at this doping are -1D like planes almost parallel to the $(k_x,k_y)$-plane, the $\delta$- FSs take the shape of six bent tubes almost perpendicular to the $(k_x,k_y)$-plane. Figure~\ref{gap} (a) - (d) show that this gap function is six-folded rotation symmetric about the $z$-axis, and changes sign upon reflection about the $k_z=0$ plane, consistent with the $p_z$-wave pairing symmetry. Besides the aspect of symmetry, Fig.~\ref{gap} (a) - (d) additionally show that the amplitude of the pairing gap on the $\delta$- FSs is lower than that on the other three FSs. This situation is more clear in Fig.~\ref{gap} (e) which shows the $k_z$ dependence of the averaged gap function on the FSs with fixed $k_z$. The reason for this lies in that the $\mathbf{k}$-dependence of the gap function of the $p_z$-wave SC in the system can be approximated as $\Delta_{\mathbf{k}}\sim \Delta_0 \sin k_z$, which is small in the small $k_z$ regime $k_z\in (-1,1)$ occupied by the $\delta$- FSs and large in the $k_z\in \pm(1.0, 1.6)$ regime occupied by the other three FSs. Similar situation is also verified for the doping level $x$ slightly higher than the Lifshitz-transition point, with only the tube-like $\delta$- FSs replaced by the 3D tube-like part of the $\gamma$- FSs, with both occupying the small $k_z$ regime $k_z \in (-1,1)$.

\section{Lifshitz-transition-enhanced $p_z$-wave SC}
 		\label{sec:Lifshitz}
 \begin{figure}[htbp]
\centering
\includegraphics[width=0.5\textwidth]{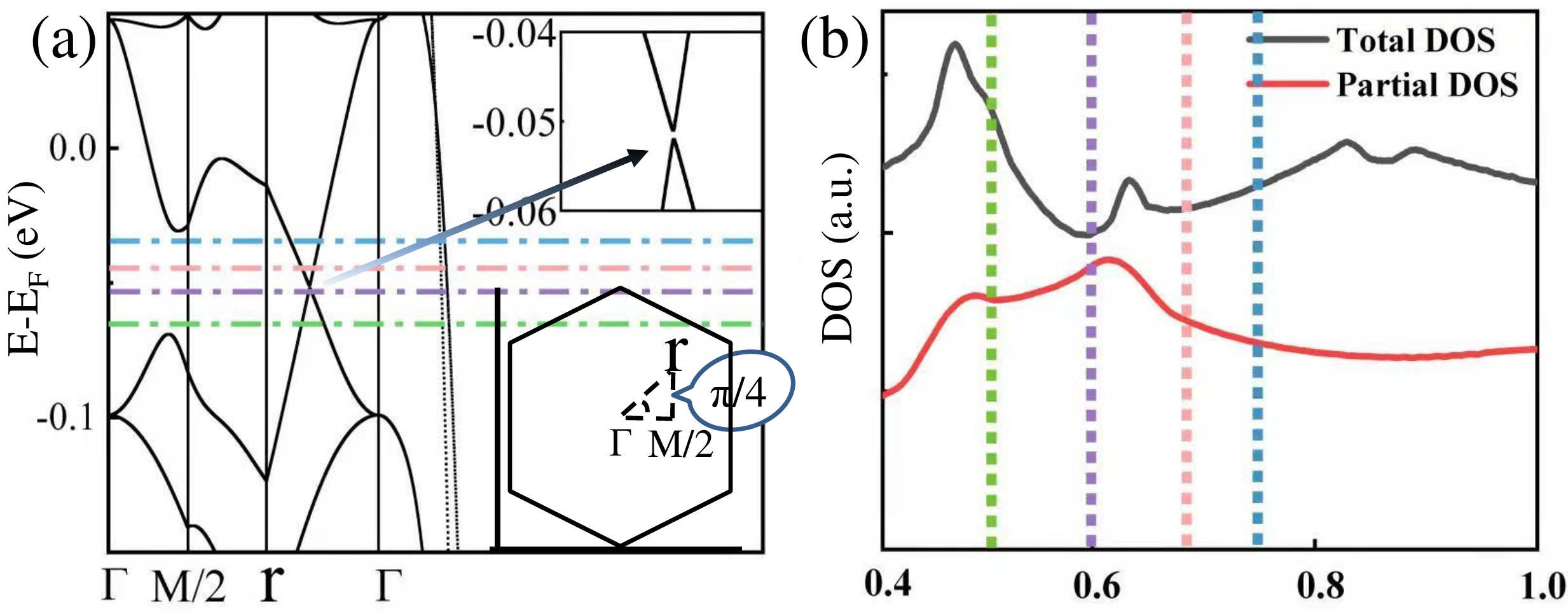}
	\caption{(color online). (a) Band structure of KCr$_3$As$_3$H along the specified path as show in the bottom inset in where $\left | \text{r}_x\right |$=$\left |\text{r}_y\right |$=$\left |\text{M}/2\right |$. The enlarged approximate ``Dirac cone'' in the band structure is shown in the upper inset. (b) The doping $x$-dependence of the total DOS (black) and the partial DOS (red) contributed by $k_z\in (1,1.6)$ on the FSs. The dash-dot lines in (a) and dotted lines in (b) with different colors mark different $x$, with the same convention as used in Fig.~\ref{chi_sc_phase}. }
 	 	\label{band2}
\end{figure}

To understand the physical origin of the triplet $p_z$-wave pairing as well as the dome-shaped $\lambda\sim x$ relation curve peaking near the Lifshitz-transition point $x=0.6$ as shown in Fig.~\ref{chi_sc_phase} (b), let's perform a more thorough investigation on the detailed band structure and the doping dependence of the FSs near the Lifshitz transition.

From Fig.~\ref{structure} (b), there is a gap between the $\gamma$- and $\delta$- bands on the $k_z=0$ plane slightly below the Fermi energy of KHCr$_3$As$_3$. The Fermi energy of the Lifshitz transition doping $x=0.6$ is just located within this gap. To more clearly reflect the low-energy band structure near this Fermi energy, in Fig.~\ref{band2} (a), we choose a specified path on the $k_z=0$ plane shown in the bottom inset, which shows that the $\gamma$- and the $\delta$- bands nearly cross each other, opening a tiny gap of about 1 meV shown in the upper inset, forming an approximate-Dirac- Fermi point at $x=0.6$. This approximate Dirac- crossing suppresses the density of state (DOS) nearby, as verified by the dip in the total-DOS curve shown in Fig.~\ref{band2} (b). It's then a puzzle why the $T_c\sim x$ relation of the $p_z$-wave SC shown in Fig.~\ref{chi_sc_phase} (b) peaks near the Lifshitz-transition doping, as the suppressed DOS there is generally harmful for the formation of SC.

The solution of this puzzle lies in a known routine which governs the distribution of the pairing gap function on the FS of an e-e interaction driven superconductor: the regimes with relatively large DOS on the FS should be distributed with relatively large pairing gap amplitudes, so that the system can gain more energy from the superconducting condensation~\cite{HuJ}. Figure~\ref{band2} (a) and our following analysis for the doping dependence of the FSs both suggest that the Lifshitz-transition mainly suppresses the DOS contributed by the small $\left|k_z\right|$ regime $k_z\in (-1,1)$. As Fig.~\ref{gap} shows that the $p_z$-wave pairing amplitude is low in the regime $k_z\in (-1,1)$, the Lifshitz transition is not harmful to the formation of SC with this pairing symmetry. More importantly, if we focus on the partial DOS contributed by the large $\left|k_z\right|$ regime $k_z\in \pm(1.0, 1.6)$ on the FSs including the $\alpha$-, $\beta$- FSs and the quasi-1D part of the $\gamma$ FSs, this part of DOS takes a peak near the Lifshitz transition doping, as shown in Fig.~\ref{band2} (b). Such a maximized DOS in these $k_z$ regimes on the FSs favors the formation of the $p_z$-wave pairing since its pairing amplitude, approximately proportional to $\sin k_z$, is large in these $k_z$ regimes. Therefore, the partial-DOS peak shown in Fig.~\ref{band2} (b), in combination with the $k_z$-dependence of the $p_z$-wave pairing gap function shown in Fig.~\ref{gap} can well account for the domed like $T_c\sim x$ relation for the $p_z$-wave SC shown in Fig.~\ref{chi_sc_phase} (b).

Two further questions arise. Why the partial DOS contributed by the large $\left|k_z\right|$ regimes is maximized near the Lifshitz transition? And why the triplet $p_z$-wave pairing is favored? The answer for the two questions lies in the presence of the type-II VHSs~\cite{VHS1,VHS2,VHS3,VHS4,VHS5} in the band structure. To clarify this point, a thorough investigation on the evolution of the FSs with $x$ is necessary. For this purpose, we choose four typical dopings $x=0.75, 0.67, 0.6, 0.5$ marked in Fig.~\ref{chi_sc_phase} with their Fermi energies marked in Fig.~\ref{band2} (a), under which the $p_z$-wave pairing dominates other pairing symmetries. In the following, we shall study the 3D FSs and typical 2D cuts of the FSs in the fixed $k_z$ planes for the four doping levels. We shall focus on the $\gamma$- and $\delta$- FSs which will experience important variation with $x$, and ignore the $\alpha$- and $\beta$- FSs.

\begin{figure*}[htbp]
	\centering
	\includegraphics[width=6in]{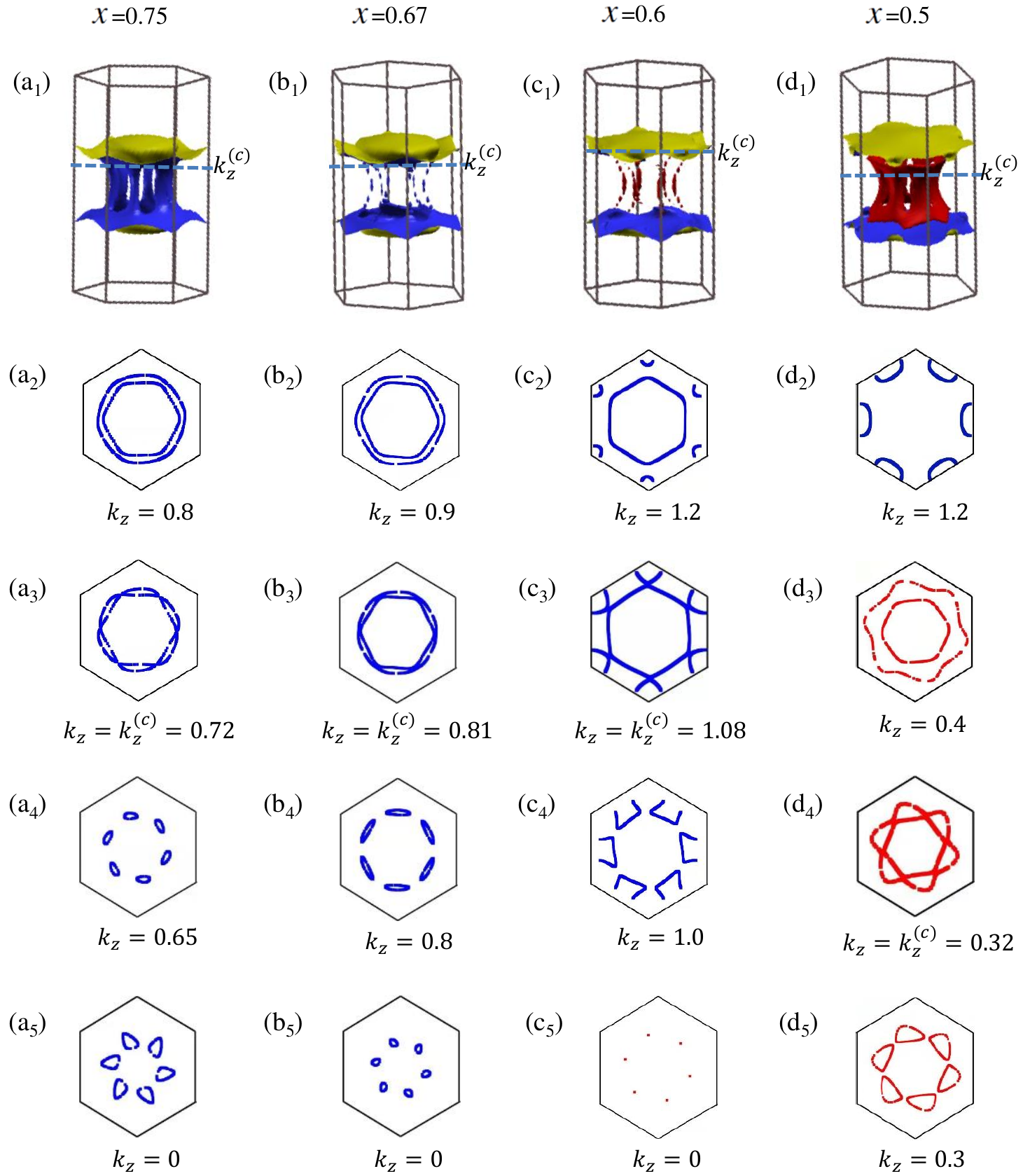}
	\caption{ (color online). 3D FSs and 2D cuts of the FSs on fixed $k_z$ planes for KCr$_3$As$_3$H$_x$ with typical doping levels $x=0.75$ in (a$_1$) - (a$_5$), $x=0.67$ in (b$_1$) - (b$_5$), $x=0.6$ in (c$_1$) - (c$_5$) and $x=0.5$ in (d$_1$) - (d$_5$) under the rigid-band approximation. For each $x$, the figure in the first row shows the 3D FSs and those in the remaining four rows show the 2D fixed-$k_z$ cuts of the FSs with different $k_z$ marked in each figure. Note that for each $x$, there exists a critical $k_z$ marked as $k_z^{(c)}$: the 2D cuts of the FSs on the $k_z=\pm k_z^{(c)}$ plane is experiencing a 2D Lifshitz transition, forming the type-II VHSs~\cite{VHS1,VHS2,VHS3,VHS4,VHS5}. The values of $k_z^{(c)}$ are marked in (a$_3$), (b$_3$), (c$_3$) and (d$_4$) for the four doping levels respectively. The position of the $k_z=k_z^{(c)}$ plane is also marked in the 3D FSs for each doping. The blue color marks the $\gamma$-FSs and the red one marks the $\delta$-FSs.}\label{fs}
\end{figure*}	

For the doping level $x=0.75$, the 3D FS is shown in Fig.~\ref{fs} (a$_1$), and the four typical 2D FS cuts are shown in Fig.~\ref{fs} (a$_2$) - (a$_5$). For $x=0.75$, the FS is only contributed from the $\gamma$- band, and the $\delta$- FS is absent. Figure~\ref{fs} (a$_1$) shows that this FS is globally connected, consisting of two flat quasi-1D sheets nearly parallel with the $(k_x,k_y)$ plane, connected by six bent tube-like FS sheets. The 3D FS cuts the $k_z=0$ plane to form six symmetry-related pockets, as shown in Fig.~\ref{fs} (a$_5$). With increasing $\left|k_z\right|$, the size of the Fermi pockets in the fixed $k_z$ cuts initially varies nonmonotonicly, and finally increases monotonicly until the adjacent pockets touch each other at the critical $k_z=\pm k_z^{(c)}=\pm 0.72$ to form a 2D Lifshitz transition, as shown in Fig.~\ref{fs} (a$_3$). The  FS cuts on the $k_z=\pm k_z^{(c)}$ planes can be viewed as the boundary between the quasi-1D FS sheets and the tube-like FS sheets, as shown by the dashed line in Fig.~\ref{fs} (a$_1$). The FS cuts for $k_z=0.8$ and $k_z=0.65$ are shown in Fig.~\ref{fs} (a$_2$) and (a$_4$) for comparison. Obviously, the 2D Lifshitz transitions at $k_z=\pm k_z^{(c)}$ form the so-called type-II VHS~\cite{VHS1,VHS2,VHS3,VHS4,VHS5}, in which the VH momenta {\bf are not} located at time-reversal variant points. It's pointed out~\cite{VHS1,VHS2,VHS3,VHS4,VHS5} that triplet SC would generally be favored near the type-II VHS, mediated by the ferromagnetic fluctuations brought about by the strong forward scatterings there. This explains the origin of the triplet SC in the system.

 With decreasing $x$, the six tube-like FS sheets become thinner, until each of them is broken into several segments below $x\approx 0.7$, as shown in Fig.~\ref{fs} (b$_1$) for $x=0.67$. Clearly, the size of the Fermi pockets on the $k_z=0$ cut shown in Fig.~\ref{fs} (b$_5$) for $x=0.67$ is smaller than that for $x=0.75$. For $x=0.67$, the six topmost and six bottommost tube-like FS sheets also grow thick enough at $k_z=\pm k_z^{(c)}=\pm 0.81$ so that any two adjacent tubes touch each other and consequently these tubes promptly evolve into the two flat quasi-1D FS sheets. The 2D FS cut at $k_z=k_z^{(c)}=0.81$ shown in Fig.~\ref{fs} (b$_3$) again illustrates the type-II VHS, with the cases for $k_z=0.8$ and $0.9$ shown for comparison. Clearly, the $k_z^{(c)}$ is enhanced at this doping. Such enhancement of the $k_z^{(c)}$ favors the formation of the $p_z$-wave pairing because the divergence of 2D DOS takes place at enhanced $\left|k_z\right|$ where the $p_z$-wave pairing gap amplitude is larger. This explains why the $T_c$ for the $p_z$-wave SC is enhanced from $x=0.75$ to $x=0.67$.

When $x$ further decreases to $x=0.6$, all the inner broken segments of the six tube-like $\gamma$- FS sheets vanish while the topmost and the bottommost segments still exist and are connected to the flat quasi-1D FS sheets, identifying the 3D-quasi-1D Lifshitz transition, as shown in Fig.~\ref{fs} (c$_1$). The residual topmost and bottommost segments are now more appropriately described as twelve thin antennas stuck out from the flat quasi-1D FS sheets. The $k_z^{(c)}$ now attains its maximum value $k_z^{(c)}=1.08$ as shown in Fig.~\ref{fs} (c$_3$), and (c$_2$) and (c$_4$) for comparison, leading to the largest $T_c$ for the $p_z$-wave pairing around $x=0.6$. Meanwhile, another six separate tube-like $\delta$- FS sheets (red colored) appear, which cuts the $k_z=0$ plane to form six very small pockets shown in Fig.~\ref{fs} (c$_5$). These six tube-like $\delta$-FSs grow thicker and thicker when $x$ further decreases, and finally each two adjacent tubes touch each other again, as shown in Fig.~\ref{fs} (d$_1$) for $x=0.5$. At $x=0.5$, although the type-II VHSs are no longer present on any 2D fixed $k_z$ cuts of the $\gamma$-FSs, as shown in Fig.~\ref{fs} (d$_2$) for a typical $k_z=1.2$, they appear on the 2D cuts of the $\delta$-FSs instead at $k_z=\pm k_z^{(c)}=\pm 0.32$, as shown in Fig.~\ref{fs} (d$_4$), and (d$_3$) and (d$_5$) for comparison. Although these $\delta$-band type-II VHSs also favor the triplet pairing, its $T_c$ is lower than that of $x=0.6$ as the $k_z^{(c)}$ is largely suppressed. Now we understand why the $T_c$ for the $p_z$-wave SC is highest around the 3D-quasi-1D Lifshitz-transition doping $x=0.6$.

To summarize this section, from detailed analysis on the hydrogen-doping $x$ dependence of the 3D FSs and 2D cuts of the FSs in the fixed $k_z$ planes, we have revealed the origin of the triplet $p_z$-wave SC as well as the dome-shaped $T_c\sim x$ relation peaking at the 3D-quasi-1D Lifshitz transition doping. It turns out that the $\gamma$-band contributes a special 3D FS which consists of two flat quasi-1D FS sheets connected by six tube-like FS sheets, the boundaries between the two parts locate within two fixed $k_z$ planes with $k_z=\pm k_z^{(c)}$. It's important that the type-II VHSs appear on these two boundaries, which favor the formation of triplet SC. What's more, the $k_z^{(c)}$ is largest near the 3D-quasi-1D Lifshitz transition doping, which pushes the $T_c$ of the triplet $p_z$-wave SC to its maximum because its $\Delta_{\mathbf{k}}\sim \Delta_0 \sin k_z$ gap form factor likes the VHSs with enhanced DOS locating at larger $\left|k_z\right|$.

\section{Discussion and Conclusion}
\label{sec:summary}

In conclusion, adopting the TB model constructed from the DFT band structure equipped with the extended Hubbard interactions, we use the RPA approach to study the pairing state of the hydrogen doped KCr$_3$As$_3$ under the rigid-band approximation. In the physically reasonable hydrogen-doping regime $x\in(0.4,1)$ where evidence of SC has been experimentally identified, our RPA results yield the triplet $p_z$-wave pairing as the leading pairing symmetry. The $T_c\sim x$ relation for the $p_z$-wave SC takes a domed shape peaking at the 3D-quasi-1D Lifshitz transition doping level. The physical origin of the triplet $p_z$-wave SC and its dome-shaped $T_c\sim x$ relation is related to the presence of the type-II VHSs~\cite{VHS1,VHS2,VHS3,VHS4,VHS5} on the $\gamma$- FS, owing to its special structure consisting of two flat quasi-1D FS sheets connected by six tube-like FS sheets, as has been summarized in the last paragraph in Sec.~\ref{sec:Lifshitz}.

Note that the Lifshitz-transition doping level in our TB model is $x_c=0.6$, which is slightly different from the $x_c=0.73$ in our DFT band structure obtained via the QE code and the $x_c=0.75$ in previous DFT band structure obtained via the VASP code~\cite{Wu:19}, due to the deviation in the TB fitting. However, the detailed band structures and the shapes of the FSs for the three are similar near their Lifshitz-transition doping levels. Therefore, the $T_c\sim x$ relation for the realistic DFT band structures should take similar domed shapes peaking near the Lifshitz-transition dopings $x_c=0.73$ or $0.75$~\cite{Xiang:20}, which are near the optimum doping $x_{\text{opt}}\in(0.65,0.71)$ estimated from experiments\cite{Taddei:17a}. Such a dome-shaped $T_c\sim x$ relation can serve as a mark to distinguish the e-e interaction-driven $p_z$-wave SC from the $s$-wave SC mediated by electron-phonon coupling, because if the pairing mechanism  is the latter, the $T_c$ should peak at the DOS maximum, while the Lifshitz-transition just takes place at the doping level of DOS minimum instead.

Here we have neglected the spin-orbit-coupling (SOC) in the system as the SOC for the Cr-3d orbitals is weak. In the absence of SOC, the three spin components of the spin-triplet $p_z$-wave pairing are exactly degenerate. To lift up this degeneracy, a weak atomic SOC~\cite{Wu:15,ZhangLD:19,TSC} adapting to the lattice symmetry can be added to the TB model. The resulting triplet-pairing component can be either $\uparrow\uparrow, \downarrow\downarrow$ with $S_z=\pm 1$ or $\uparrow\downarrow+\downarrow\uparrow$ with $S_z=0$. If the latter is favored, the pairing state of the system would be an spin-U(1)-symmetry protected topological SC similar with K$_2$Cr$_3$As$_3$~\cite{TSC}, hosting exactly flat surface bands on the $\left(0,0,1\right)$ surface, which can serve as a smoking-gun evidence for the $p_z$-wave SC.

 		\section*{Acknowledgements}
 We are grateful to the stimulating discussions with G.-H Cao. This work is supported by the NSFC under the grant NO.12074031 and No.11674025.

\newpage
\appendix

\begin{widetext}
	\section{Appendix: The multi-orbital RPA approach}	
	
	The Hamitonian adopted in our calculations is
	\begin{eqnarray}
	H&=&H_{\rm TB}+H_{int}\nonumber\\
	H_{int}&=&U\sum_{i\mu}n_{i\mu\uparrow}n_{i\mu\downarrow}+
	V\sum_{i,\mu<\nu}n_{i\mu}n_{i\nu}+J_{H}\sum_{i,\mu<\nu}\Big[
	\sum_{\sigma\sigma^{\prime}}c^{+}_{i\mu\sigma}c^{+}_{i\nu\sigma^{\prime}}
	c_{i\mu\sigma^{\prime}}c_{i\nu\sigma}+(c^{+}_{i\mu\uparrow}c^{+}_{i\mu\downarrow}
	c_{i\nu\downarrow}c_{i\nu\uparrow}+h.c.)\Big]\label{H-H-model}
	\end{eqnarray}
	Let's define the following bare susceptibility for the non-interacting case ($U=V=J_H=0$),
	\begin{equation}
	\chi^{(0)l_{1},l_{2}}_{l_{3},l_{4}}\left(\mathbf{q},\tau\right)\equiv
	\frac{1}{N}\sum_{\mathbf{k_{1},k_{2}}}\left<T_{\tau}c^{\dagger}_{l_{1}}(\mathbf{k_{1}},\tau)
	c_{l_{2}}(\mathbf{k_{1}+q},\tau)c^{+}_{l_{3}}(\mathbf{k_{2}+q},0)c_{l_{4}}(\mathbf{k_{2}},0)\right>_0,\label{bare}
	\end{equation}
	where $l_{i}$ $(i=1,\cdots,4)$ denote orbital indices. The explicit formulism of $\chi^{(0)}$ in the momentum-frequency space is,
	\begin{equation}
	\chi^{(0)l_{1},l_{2}}_{l_{3},l_{4}}\left(\mathbf{q},i\omega_n\right)=\frac{1}{N}\sum_{\mathbf{k},\alpha,\beta}
	\xi_{l_{1}}^{\alpha,*}(\mathbf{k})\xi^{\beta}_{l_{2}}(\mathbf{k+q})\xi^{\beta,*}_{l_{3}}(\mathbf{k+q})\xi^{\alpha}_{l_{4}}(\mathbf{k})
	\frac{n_{F}(\varepsilon^{\beta}_{\mathbf{k+q}})-n_{F}(\varepsilon^{\alpha}_{\mathbf{k}})}{i\omega_n+\varepsilon^{\alpha}_{\mathbf{k}}-\varepsilon^{\beta}_{\mathbf{k+q}}},\label{explicit_free}
	\end{equation}
	where $\alpha/\beta=1,...,6$ are band indices, $\varepsilon^{\alpha}_{\mathbf{k}}$ and $\xi^{\alpha}_{l}\left(\mathbf{k}\right)$ are the $\alpha-$th eigenvalue and eigenvector
	of the $H_{TB}(\mathbf{k})$ matrix respectively and $n_F$ is the Fermi-Dirac distribution function.
	
	When the Hubbard interaction in Eq. (\ref{H-H-model}) is included, we can explicitly calculate the spin ($\chi^{(s)}$) and charge ($\chi^{(c)}$) susceptibilities as follow,
	\begin{eqnarray}
	\chi^{(c)l_{1},l_{2}}_{l_{3},l_{4}}\left(\mathbf{q},\tau\right)&\equiv&
	\frac{1}{2N}\sum_{\mathbf{k_{1},k_{2}},\sigma_{1},\sigma_{2}}\left<T_{\tau}C^{\dagger}_{l_{1},\sigma_{1}}(\mathbf{k_{1}},\tau)
	C_{l_{2},\sigma_{1}}(\mathbf{k_{1}+q},\tau)C^{+}_{l_{3},\sigma_{2}}(\mathbf{k_{2}+q},0)C_{l_{4},\sigma_{2}}(\mathbf{k_{2}},0)\right>,
	\nonumber\\
	\chi^{(s)l_{1},l_{2}}_{l_{3},l_{4}}\left(\mathbf{q},\tau\right)&\equiv&
	\frac{1}{2N}\sum_{\mathbf{k_{1},k_{2}},\sigma_{1},\sigma_{2}}\sigma_{1}\sigma_{2}\left<T_{\tau}C^{\dagger}_{l_{1},\sigma_{1}}(\mathbf{k_{1}},\tau)
	C_{l_{2},\sigma_{1}}(\mathbf{k_{1}+q},\tau)C^{+}_{l_{3},\sigma_{2}}(\mathbf{k_{2}+q},0)C_{l_{4},\sigma_{2}}(\mathbf{k_{2}},0)\right>.
	\end{eqnarray}

	Note that when $U=V=J_H=0$ we have $\chi^{(c)}=\chi^{(s)}=\chi^{(0)}$.
	In the RPA level, the Cooper pair with momentum and orbital of $(\bm{k}l_3,-\bm{k}l_4)$ could be scattered into $(\bm{k'}l_1,-\bm{k'}l_2)$ by exchanging charge or spin fluctuations. This process can be explained graphically by Feynman diagrams shown as Fig.~(\ref{feymman1}).
	
	\begin{figure}[htbp]
		\centering
		\includegraphics[width=0.49\textwidth]{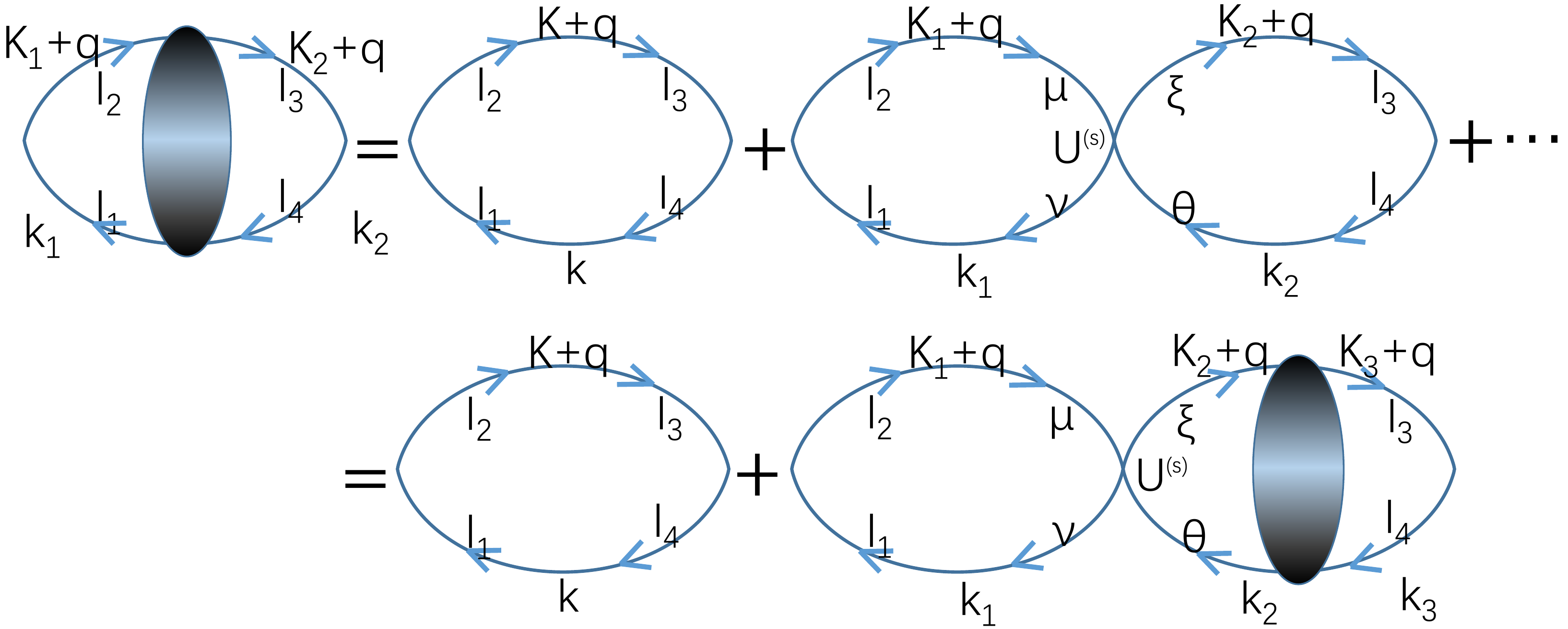}
		\caption{Feynman's diagram for the renormalized susceptibilities
			in the RPA level.}\label{feymman1}
	\end{figure}
	
	The renormalized spin/charge susceptibilities for the system are,
	\begin{align}\label{chisc}
	\chi ^{(s/c)}(\bm{k},i\omega _{n})=[I\mp\chi ^{(0)}(\bm{k},i\omega _{n})U^{s/c}]^{-1}\chi ^{(0)}(\bm{k},iw_{n}).		
	\end{align}
	where $\chi^{(s,c)}\left(\mathbf{k},i\omega _{n}\right)$, $\chi^{(0)}\left(\mathbf{k},i\omega _{n}\right)$ and $U^{(s,c)}$ are operated as
	$36\times 36$ matrices (the upper or lower two indices are viewed as one number), the nonzero elements $U^{(s/c)l_{1}l_{2}}_{l_{3}l_{4}}$ of $U^{s/c}$ are as follows,
	\begin{eqnarray}
	U^{(s) l_{1}l_{2}}_{l_{3}l_{4}}&=&\left\{\begin{array}{cc}{U,l_{1}=l_{2}=l_{3}=l_{4}}\\{J_H,l_{1}=l_{2}\ne
		l_{3}=l_{4}}\\{J_{H},l_{1}=l_{3}\ne
		l_{2}=l_{4}}\\{V,l_{1}=l_{4}\ne
		l_{3}=l_{2}}\end{array}\right.
	\end{eqnarray}
	
	\begin{eqnarray}
	U^{(c) l_{1}l_{2}}_{l_{3}l_{4}}&=&\left\{\begin{array}{cc}{U,l_{1}=l_{2}=l_{3}=l_{4}}\\{2V-J_H,l_{1}=l_{2}\ne
		l_{3}=l_{4}}\\{J_{H},l_{1}=l_{3}\ne
		l_{2}=l_{4}}\\{2J_{H}-V,l_{1}=l_{4}\ne
		l_{3}=l_{2}}\end{array}\right.
	\end{eqnarray}
	
	For repulsive Hubbard-interactions, the spin susceptibility is enhanced and the charge susceptibility is suppressed. Note that there is a critical interaction strength $U_c$ which depends on the ratio $J_H/U$. Note that when the interaction strength $U$ is higher than $U_c$, the denominator matrix $I-\chi ^{(0)}(\bm{k},iw_{n})U^{s}$ in Eq. (\ref{chisc}) will have zero eigenvalues for some $\bm q$ and the renormalized spin susceptibility diverges there, which invalidates the RPA treatment. When $U<U_c$, the short-ranged spin or charge fluctutions would mediate Cooper pairing in the system.
	
	Considering a Cooper pair with momentum/orbital $(\bm k't,-\bm k's)$, it could be scattered to $(\bm kp,-\bm kq)$ by exchanging charge or spin fluctuations. In the RPA level, The effective interaction induced by this process is as follows:

	\begin{align}\label{veff}
	\mathrm{V}_{eff}^{\mathrm{RPA}}=\frac{1}{\mathrm{N}}\sum_{pqst,\bm k\bm k'}\Gamma _{pq}^{st}(\bm k,\bm k')c_{p}^{+}(\bm k)c_{q}^{+}(-\bm k)c_{s}(-\bm k')c_{t}(\bm k').
	\end{align}
	
	\begin{figure}[htbp]
		\centering
		\includegraphics[width=0.49\textwidth]{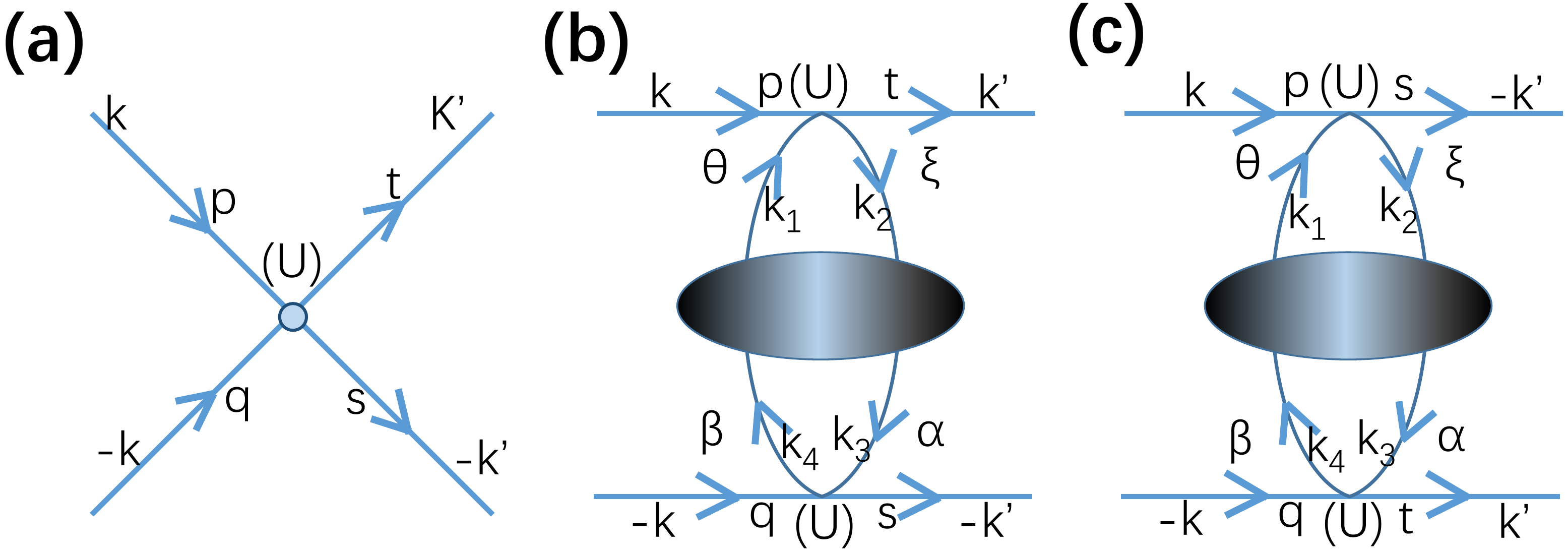}
		\caption{Three processes which contribute the renormalized effective vertex considered in the RPA, with (a) the bare interaction vertex and
			(b),(c) the two second order perturbation processes during which spin or charge fluctuations are exchanged between a Cooper pair.}\label{feymman2}
	\end{figure}

	We consider the three processes in Fig. \ref{feymman2} \cite{Ming} which contribute
	to the effective vertex $\Gamma_{st}^{pq}(\bm k,\bm k')$, where (a) represents the bare interaction vertex and (b),(c) represent the two second order perturbation processes during which spin or charge fluctuations are exchanged between a Cooper pair. Hence this effective interaction process can be divided by spin pairings into singlet channel and triplet channel.
	
	In the singlet channel, the effective vertex $\Gamma^{pq}_{st}(k,k')$ is given as follow,
	\begin{eqnarray}
	\Gamma^{pq(s)}_{st}(k,k')=\left(\frac{U^{(c)}+3U^{(s)}}{4}\right)^{pt}_{qs}+
	\frac{1}{4}\left[3U^{(s)}\chi^{(s)}\left(k-k'\right)U^{(s)}-U^{(c)}\chi^{(c)}\left(k-k'\right)U^{(c)}\right]^{pt}_{qs}+\nonumber\\
	\frac{1}{4}\left[3U^{(s)}\chi^{(s)}\left(k+k'\right)U^{(s)}-U^{(c)}\chi^{(c)}\left(k+k'\right)U^{(c)}\right]^{ps}_{qt},
	\end{eqnarray}
	while in the triplet channel, it is
	\begin{eqnarray}
	\Gamma^{pq(t)}_{st}(k,k')=\left(\frac{U^{(c)}-U^{(s)}}{4}\right)^{pt}_{qs}-
	\frac{1}{4}\left[U^{(s)}\chi^{(s)}\left(k-k'\right)U^{(s)}+U^{(c)}\chi^{(c)}\left(k-k'\right)U^{(c)}\right]^{pt}_{qs}+\nonumber\\
	\frac{1}{4}\left[U^{(s)}\chi^{(s)}\left(k+k'\right)U^{(s)}+U^{(c)}\chi^{(c)}\left(k+k'\right)U^{(c)}\right]^{ps}_{qt},
	\end{eqnarray}

	Notice that the vertex $\Gamma^{pq}_{st}(k,k')$ has been symmetrized for the singlet case and anti-symmetrized for the
	triplet case. Generally we neglect the frequency-dependence of $\Gamma$ and replace it by $\Gamma^{pq}_{st}(k,k')\approx\Gamma^{pq}_{st}(\mathbf{k,k'},0)$.
	
	Considering only intra-band pairings, we obtain the following effective pairing interaction on the FS,
	\begin{eqnarray}
	V_{eff}=
	\frac{1}{N}\sum_{\alpha\beta,\mathbf{k}\mathbf{k'}}V^{\alpha\beta}(\mathbf{k,k'})c_{\alpha}^{\dagger}(\mathbf{k})
	c_{\alpha}^{\dagger}(-\mathbf{k})c_{\beta}(-\mathbf{k}')c_{\beta}(\mathbf{k}').\label{pairing_interaction}
	\end{eqnarray}
	where $\alpha/\beta=1,\cdots,6$ are band indices and the energy gap equation
	\begin{align}\label{gap1}
	\Delta_{\boldsymbol{k}}^{\alpha}=\frac{1}{N} \sum_{\boldsymbol{k}^{\prime}, \beta} V^{\alpha \beta}\left(\boldsymbol{k}, \boldsymbol{k}^{\prime}\right)\left\langle c_{\beta \downarrow}\left(-\boldsymbol{k}^{\prime}\right) c_{\beta \uparrow}\left(\boldsymbol{k}^{\prime}\right)\right\rangle.
	\end{align}
	
	The Hamiltonian in Eq.~\ref{H-H-model} becomes
	\begin{align}
	H_{mf}=\sum_{\boldsymbol{k}, \alpha, \sigma} \varepsilon_{\boldsymbol{k}}^{\alpha} c_{\alpha \sigma}^{\dagger}(\boldsymbol{k}) c_{\alpha \sigma}(\boldsymbol{k})+\sum_{\boldsymbol{k}, \alpha}\left(\Delta_{\boldsymbol{k}}^{\alpha} c_{\alpha \uparrow}^{\dagger}(\boldsymbol{k}) c_{\alpha \downarrow}^{\dagger}(-\boldsymbol{k})+h . c .\right)
	\end{align}
	Under the Bogliubov transformation
	\begin{align}
	c_{\alpha \uparrow}\left(\boldsymbol{k}\right)&=u_\alpha\left(\boldsymbol{k}\right)\gamma_{\alpha \uparrow}\left(\boldsymbol{k}\right)+v_\alpha\left(\boldsymbol{k}\right)\gamma_{ \alpha\downarrow}^{\dagger}\left(-\boldsymbol{k}\right), \\\nonumber
	c_{\alpha \downarrow}\left(-\boldsymbol{k}\right)&=u_\alpha\left(\boldsymbol{k}\right)\gamma_{\alpha \downarrow}\left(-\boldsymbol{k}\right)-v_\alpha\left(\boldsymbol{k}\right)\gamma_{ \alpha\uparrow}^{\dagger}\left(-\boldsymbol{k}\right), \\ \nonumber
	&u_\alpha^2\left(\boldsymbol{k}\right)=\frac{1}{2}(1+\frac{\varepsilon^{\alpha}_{\boldsymbol{k}}}{\xi^{\alpha}_{\boldsymbol{k}}}),\\\nonumber
	&v_\alpha^2\left(\boldsymbol{k}\right)=\frac{1}{2}(1-\frac{\varepsilon^{\alpha}_{\boldsymbol{k}}}{\xi^{\alpha}_{\boldsymbol{k}}}), \\\nonumber
	&\xi^{\alpha}_{\boldsymbol{k}}=\sqrt{(\varepsilon^{\alpha}_{\boldsymbol{k}})^2
		+\left|\Delta^{\alpha}_{\boldsymbol{k}}\right|^{2}}
	\end{align}
	the mean-field Hamiltonian becomes diagonal, and the gap equation becomes
	
	\begin{align}
	\Delta_{\boldsymbol{k}}^{\alpha}&=-\frac{1}{N} \sum_{\boldsymbol{k}^{\prime}, \beta} V^{\alpha \beta}\left(\boldsymbol{k}, \boldsymbol{k}^{\prime}\right) \frac{\Delta_{\boldsymbol{k}^{\prime}}^{\beta}}{2 \xi_{\boldsymbol{k}^{\prime}}}\left[1-2f(\xi_{\boldsymbol{k}^{\prime}})\right]\\\nonumber
	&=-\frac{1}{(2 \pi)^{2}} \sum_{\beta} \int d k_{\|}^{\prime} \int d k_{\perp}^{\prime} V^{\alpha \beta}\left(\boldsymbol{k}, \boldsymbol{k}^{\prime}\right) \frac{\Delta_{\boldsymbol{k}^{\prime}}^{\beta}}{2\xi_{\boldsymbol{k}^{\prime}}^{\beta}} \tanh \left(\frac{\xi_{\boldsymbol{k}^{\prime}}^{\beta}}{2 k_{B} T}\right).
	\end{align}
	It is noted that the main contribution to the above integration comes from the momenta near the Fermi surface, where $\varepsilon^{\alpha}_{\boldsymbol{k}}=\nu^{\alpha}_{F}({\boldsymbol{k}})k_{\perp}$.
	Near the superconducting critical temperature $T_c$, $\Delta_{\boldsymbol{k}}^{\alpha}$ tends to be zero. Up to the first-order term of $\Delta_{\boldsymbol{k}}^{\alpha}$, the Eq.~\ref{gap1} becomes the following linearized one\cite{Scalapino2009,Scalapino2011,Wu:15}:
	\begin{eqnarray}
	V^{\alpha\beta}(\mathbf{k,k'})=\sum_{pqst,\mathbf{k}\mathbf{k'}}\Gamma^{pq}_{st}(\mathbf{k,k'},0)\xi_{p}^{\alpha,*}(\mathbf{k})
	\xi_{q}^{\alpha,*}(-\mathbf{k})\xi_{s}^{\beta}(-\mathbf{k'})\xi_{t}^{\beta}(\mathbf{k'}).
	\end{eqnarray}
	From the effective pairing interaction (\ref{pairing_interaction}), one can obtain the following linearized gap
	equation~\cite{Scalapino2009,Scalapino2011,Wu:15} to determine the $T_c$ and the leading pairing symmetry of the system,
	\begin{equation}
	-\frac{1}{(2\pi)^3}\sum_{\beta}\oiint_{FS}
	d^{2}\mathbf{k'}_{\Vert}\frac{V^{\alpha\beta}(\mathbf{k,k'})}{v^{\beta}_{F}(\mathbf{k'})}\Delta_{\beta}(\mathbf{k'})=\lambda
	\Delta_{\alpha}(\mathbf{k}).\label{eigenvalue_Tc}
	\end{equation}
	This equation can be looked upon as an eigenvalue problem, where the normalized eigenvector $\Delta_{\alpha}(\mathbf{k})$ represents the relative gap function on the $\alpha-$th FS patches near $T_c$, and eigenvalue $\lambda$ is related to $T_c$ through $\lambda^{-1}=\ln \left(1.13 \frac{\hbar \omega_{D}}{k_{B} T_{c}}\right)$. The leading pairing symmetry is determined by the largest eigenvalue $\lambda$ of Eq.~(\ref{eigenvalue_Tc}).
	
\end{widetext}

\end{document}